\documentclass[%
	reprint,
	superscriptaddress,
	amsmath,amssymb,
	aps,
	pra,
]%
{revtex4-2}

\usepackage[utf8]{inputenc}
\usepackage[T1]{fontenc}
\usepackage{lmodern}
\usepackage[english]{babel}
\usepackage[autostyle=true]{csquotes}

\usepackage{hyperref}
\hypersetup{
		pdftitle={Superfluidity and sound propagation in disordered Bose gases},
		pdfauthor={Kevin T. Geier, Jeff Maki, Alberto Biella, Franco Dalfovo, Stefano Giorgini, Sandro Stringari},
}

\usepackage[dvipsnames]{xcolor}
\usepackage{graphicx}
\graphicspath{{figures/}}

\usepackage[%
	caption=false,
	subrefformat=simple,
	labelformat=simple,
]{subfig}

\usepackage{bm}
\usepackage{mathtools}
\usepackage{etoolbox}
\usepackage{dsfont}
\usepackage{xspace}

\usepackage{mleftright}
\mleftright

\usepackage[%
    separate-uncertainty=false,
    range-phrase=-,
]{siunitx}

\DeclareSIUnit\bohr{\text {\ensuremath {a}}_{0}}

\usepackage[
	capitalise,
]{cleveref}

\newcommand*{\vect}[1]{\bm{#1}}
\newcommand*{\diff}{\mathop{}\!\mathrm{d}}
\newcommand*{\etothepowerof}[1]{\mathrm{e}^{#1}}
\newcommand*{\etothe}[1]{\etothepowerof{#1}}

\DeclarePairedDelimiterX{\commutator}[2]{[}{]}{#1,#2}
\DeclarePairedDelimiterX{\anticommutator}[2]{\{}{\}}{#1,#2}

\DeclarePairedDelimiter{\abs}{\lvert}{\rvert}

\DeclareMathOperator{\real}{Re}
\DeclareMathOperator{\imag}{Im}
\newcommand*{\bigO}{\mathcal{O}}

\DeclarePairedDelimiter{\ket}{|}{\rangle}

\DeclarePairedDelimiter{\braket}{\langle}{\rangle}
\DeclarePairedDelimiterX{\ketbra}[2]{|}{|}{#1\delimsize\rangle\delimsize\langle#2}
\DeclarePairedDelimiter{\quantumaverage}{\langle}{\rangle}
\DeclarePairedDelimiterXPP\quantumequilibrium[1]{}\langle\rangle{_0}{\ifblank{#1}{\:\cdot\:}{#1}}
\DeclarePairedDelimiterXPP\quantumperturbed[1]{}\langle\rangle{_{\lambda}}{\ifblank{#1}{\:\cdot\:}{#1}}
\newcommand*{\ensembleaverage}[1]{\overline{#1}}
\DeclareMathOperator{\sinc}{sinc}
\DeclareMathOperator{\tri}{tri}

\DeclarePairedDelimiterX{\set}[1]{\{}{\}}{%
	
	#1
}

\DeclarePairedDelimiterX\norm[1]\lVert\rVert{\ifblank{#1}{\:\cdot\:}{#1}}
\DeclarePairedDelimiterXPP\twonorm[1]{}\lVert\rVert{_2}{\ifblank{#1}{\:\cdot\:}{#1}}
\DeclarePairedDelimiterXPP\infinitynorm[1]{}\lVert\rVert{_{\infty}}{\ifblank{#1}{\:\cdot\:}{#1}}

\renewcommand{\vec}{\vect}

\newcommand*{\rubidium}{$^{87}$Rb\xspace}

\begin{document}

\title{Superfluidity and sound propagation in disordered Bose gases}

\author{Kevin T. Geier}
\email[]{kevin.geier@tii.ae}
\affiliation{Pitaevskii BEC Center, CNR-INO and Dipartimento di Fisica, Universit\`a di Trento, 38123 Trento, Italy}
\affiliation{Trento Institute for Fundamental Physics and Applications, INFN, 38123 Trento, Italy}
\affiliation{Quantum Research Center, Technology Innovation Institute, P.O. Box 9639, Abu Dhabi, United Arab Emirates}

\author{Jeff Maki}
\email[]{jeffrey.maki@ino.cnr.it}
\affiliation{Pitaevskii BEC Center, CNR-INO and Dipartimento di Fisica, Universit\`a di Trento, 38123 Trento, Italy}

\author{Alberto Biella}
\affiliation{Pitaevskii BEC Center, CNR-INO and Dipartimento di Fisica, Universit\`a di Trento, 38123 Trento, Italy}

\author{Franco Dalfovo}
\affiliation{Pitaevskii BEC Center, CNR-INO and Dipartimento di Fisica, Universit\`a di Trento, 38123 Trento, Italy}

\author{Stefano Giorgini}
\affiliation{Pitaevskii BEC Center, CNR-INO and Dipartimento di Fisica, Universit\`a di Trento, 38123 Trento, Italy}

\author{Sandro Stringari}
\affiliation{Pitaevskii BEC Center, CNR-INO and Dipartimento di Fisica, Universit\`a di Trento, 38123 Trento, Italy}

\date{\today}

\begin{abstract}
Superfluidity describes the ability of quantum matter to flow without friction.
Due to its fundamental role in many transport phenomena, it is crucial to understand the robustness of superfluid properties to external perturbations.
Here, we theoretically study the effects of speckle disorder on the propagation of 
sound waves in a two-dimensional Bose--Einstein condensate at zero temperature.
We numerically solve the Gross--Pitaevskii equation in the presence of disorder and employ a superfluid hydrodynamic approach to elucidate the role of the compressibility and superfluid fraction on the propagation of sound. 
A key result is that disorder reduces the superfluid fraction and hence the speed of sound; it also introduces damping and mode coupling.
In the limit of weak disorder, the predictions for the speed of sound and its damping rate are well reproduced by a quadratic perturbation theory.
The hydrodynamic description is valid over a wide range of parameters, while discrepancies become evident if the disorder becomes too strong, the effect being more significant for disorder applied in only one spatial direction.
Our predictions are well within the reach of state-of-the-art cold-atom experiments and carry over to more general disorder potentials.
\end{abstract}


\maketitle

\begin{figure*}[htb]
	\includegraphics[width=\linewidth]{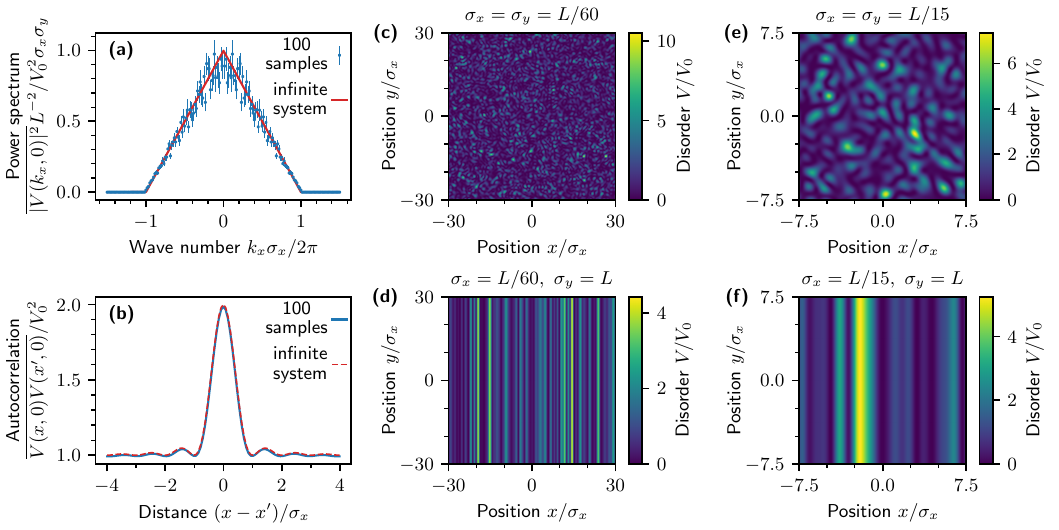}%
    \subfloat{\label{fig:speckle_disorder:a}}%
    \subfloat{\label{fig:speckle_disorder:b}}%
    \subfloat{\label{fig:speckle_disorder:c}}%
    \subfloat{\label{fig:speckle_disorder:d}}%
    \subfloat{\label{fig:speckle_disorder:e}}%
    \subfloat{\label{fig:speckle_disorder:f}}%
	\caption{\label{fig:speckle_disorder}%
        Characteristics of the speckle disorder potential.
        (a)~Disorder power spectrum, averaged over $100$ realizations, and (b) autocorrelation function for a ratio of the disorder correlation length to the box size of $\sigma_x / L = 1 / 60$.
        The error bars correspond to the standard error of the mean and the red lines show the analytical result for an infinite system according to \cref{eq:C,eq:C_x}.
        (c)--(f)~Single realizations of speckle disorder for different configurations of the correlation lengths.
        Panels (c) and (e) show isotropic 2D disorder ($\sigma_x = \sigma_y$) for $\sigma_x / L = 1/60$ and $\sigma_x / L = 1/15$, respectively, while panels (d) and (f) correspond to stripe-like 1D disorder ($\sigma_y = L$) for the same two values of $\sigma_x$.%
    }
\end{figure*}

\section{Introduction}
\label{sec:intro}

The propagation of sound in superfluid systems exhibits peculiar features which have been the subject of many fundamental theoretical and experimental investigations \cite{pitaevskii2016bose}. 
Sound can propagate in a superfluid even at zero temperature, despite the absence of thermal collisions. As outlined in the pioneering work by Bogoliubov, sound propagation in these conditions is due to fluctuations of the phase of the order parameter. 
At finite temperature, superfluidity reveals new features associated with the propagation of second sound, as predicted by  Landau's theory of two-fluid hydrodynamics \cite{pitaevskii2016bose}. 

The realization of Bose--Einstein condensation in ultracold atomic gases has provided new opportunities to confirm and explore the theoretical predictions of superfluidity. An interesting focus has been on collective oscillations and the propagation of sound in superfluid systems with broken translational invariance;
this includes the case where the invariance is broken by external fields \cite{PhysRevLett.89.170402, Kraemer2003,Zwerger2003, Taylor2003,Menotti2004, Cataliotti2001,PhysRevLett.90.140405,Chauveau2023} as well as the case where it is broken spontaneously, giving rise to supersolidity~\cite{PhysRevLett.108.175301, PhysRevLett.110.235302, Leonard2017, Li2017, PhysRevA.99.041601, Tanzi2019, Guo2019, Natale2019}. In the first case, a single class of Goldstone modes is expected to occur at zero temperature, taking the form of sound waves,  whose velocity~$c$ is predicted by the hydrodynamic theory of superfluids according to the law
\begin{equation}
mc^2=f_s (n \kappa)^{-1} \, .
\label{cHD}
\end{equation}
Here, $\kappa$ is the compressibility and
$f_s = n_s/n$ is the superfluid fraction of the system, while $n$ and $n_s$ denote the total and superfluid densities, respectively. 
The inverse compressibility is related to the chemical potential $\mu$ by the relation $\kappa^{-1}=n^2\partial \mu/\partial n$, while the superfluid fraction is related to the effective hydrodynamic mass~$m^*$ by  $f_s=m/m^*$~\cite{Chauveau2023}.
When external fields are present, due to the absence of translational invariance, $f_s$ can be significantly smaller than unity even at zero temperature, as revealed by the celebrated superfluid-to-Mott insulator transition~\cite{Greiner2002}.

In recent works, the superfluid fraction~$f_s$ has been measured in a dilute Bose--Einstein-condensed gas with density modulations created by an external optical-lattice potential~\cite{Chauveau2023,Tao2023}.
The value of $f_s$ has been obtained through the measurement of the sound velocity as a function of the intensity of the external potential, in excellent agreement with the predictions of Gross--Pitaevskii theory.
In the present paper, we investigate the propagation of sound, the role played by the superfluid fraction, and the applicability of the hydrodynamic relation~\labelcref{cHD} for weakly interacting two-dimensional Bose gases in the presence of a disordered potential.

The study of quantum fluids in disordered potentials has a long history, dating back to the dirty boson problem in condensed matter theory~\cite{Giamarchi_1987,PhysRevB.40.546} and experiments on superfluid helium adsorbed in porous glasses~\cite{PhysRevB.55.12620,PhysRevLett.84.2060}. In the context of ultracold atomic gases, experiments can be performed with random yet correlated potentials with different properties, and one can also tune the strength of the atom--atom interaction. For these reasons, ultracold atoms have proven to be an excellent platform for the study of Anderson localization~\cite{Billy2008,Roati2008,White2020} and more generally the interplay between disorder, quantum degeneracy, and interactions~\cite{PhysRevLett.95.070401, Pilati2010, Navez2006, Fallan2007, Pasienski2010, Meldgin2016}.
On the theory side, many studies have addressed the effect of disorder on the dispersion of elementary excitations in a dilute Bose gas and in particular on the propagation and damping of sound~\cite{Giorgini1994, Yukalov2007, PhysRevA.80.053620, Gaul2011, Abdullaev2012}. 
However, many of these approaches are limited to the perturbative regime where the effects of disorder are weak and the dynamics can still be accurately modelled using hydrodynamic theory. A systematic study of the effects of disorder on the sound mode at larger disorder strengths is missing from the literature to the best of our knowledge. In particular, it is unclear whether hydrodynamics is still valid in this regime, or if the sound mode is entirely destroyed by the disorder.

In this article, we perform numerical calculations based on the Gross--Pitaevskii equation (GPE) of both static and dynamic properties of a two-dimensional dilute Bose gas in a random potential modelling optical speckles. The compressibility and superfluid density are evaluated from the ground-state energy as a function of the disorder strength, for different ratios of the disorder correlation length to the healing length.
Using hydrodynamic theory, these results provide us with a determination of the speed of sound, which we compare with time-dependent simulations of sound propagation following the removal of a wave-vector-specific perturbation in the linear-response regime.
We find good agreement between the superfluid hydrodynamic theory, based on the relation~\labelcref{cHD}, and dynamic calculations for a wide parameter range well beyond the weak-disorder regime. From the simulations, we also extract the collisionless damping rate of the phonon collective mode once a statistical average over many disorder realizations is taken. The calculated decay rate agrees with perturbation theory in the weak disorder regime.

A central point of our analysis concerns the comparison between isotropic two-dimensional (2D) disorder and stripe-like one-dimensional (1D) disorder.
In the 2D case, disorder has an overall weaker impact since there are many paths for the superfluid to move around obstacles, whereas in the 1D setting, potential barriers spanning the entire system diminish the superfluid fraction already at significantly smaller disorder strengths. 
As a consequence, we find that the Leggett bounds for the superfluid fraction~\cite{Leggett1970, Leggett1998} provide too loose of an estimate for $f_s$ in the case of 2D disorder. On the contrary, in the 1D case, the upper and lower Leggett bound coincide.

Our results point toward a realistic possibility of observing these effects in current experiments, similarly to the measurement of the speed of sound in periodic potentials~\cite{Chauveau2023}.

The paper is structured as follows. In \cref{sec:Quant_dis}, we discuss the implementation of the disorder potential used in this study. \Cref{sec:TM} puts forward the theoretical methods employed, namely equilibrium simulations and time-dependent simulations of the Gross--Pitaevskii equation, and perturbation theory. The results for sound propagation and superfluid fraction for two-dimensional isotropic disorder are presented in \cref{sec:2D_disorder}, while similar results for one-dimensional stripe-like disorder are in \cref{sec:1D_disorder}. Finally, we present our conclusions in \cref{sec:conclusions}.
\Cref{app:disorder_protocol,app:PT,app:damping,app:finite_size,app:infrared_cutoff} provide further technical details for the interested reader.


\section{Speckle disorder}
\label{sec:Quant_dis}

In this work, we focus on correlated speckle disorder. Such disorder can readily be achieved in experiments by passing laser light through a diffuser~\cite{Dainty1975}. The resulting interference fringes act as a random potential imprinted onto the quantum gas.

Such speckle potentials are on average constant,
\begin{equation}
    \ensembleaverage{V(\vec{r})} = V_0 \,,
\end{equation}
while the strength of the disorder at two points is correlated,
\begin{equation}
    \ensembleaverage{[V(\vec{r})-V_0][V(\vec{r}^\prime)-V_0]} = V_0^2  C(\vec{r}-\vec{r}^\prime) \,,
\end{equation}
where $C(\vec{r}-\vec{r}^\prime)$ is the correlation function of the disorder. We use the notation $\ensembleaverage{O}$ to denote the average of an observable $O$ over many realizations of the disorder.
As outlined in \cref{app:disorder_protocol}, it is straightforward to sample such a potential numerically by writing it in terms of normally distributed random Fourier modes $a_{\vec{k}}$ with zero mean and suitably chosen variance,
\begin{equation}
    V(\vec{r}) = V_0 \, \Big|\sum_{\abs{k_i} \le \pi / \sigma_i} a_{\vec{k}} \etothe{i \vec{k} \cdot \vec{r}} \Big|^2 \,.
    \label{eq:V_def}
\end{equation}
The correlation lengths~$\sigma_i$ set the length scale of the disorder correlation function in the $i = x,y$ directions. 

In our simulations, we consider a two-dimensional gas in a $L \times L$ square box with periodic boundary conditions. Hence, the wave vectors entering the disorder obey the discretization rule  $\vec{k} = \mathinner{\vec{n}} 2 \pi / L$ with $\vec{n} \in \mathbb{Z}^2$.
In the limit $L \to \infty$, the disorder autocorrelation function is given by
\begin{subequations}
    \label{eq:C}
\begin{gather}
    \label{eq:C:a}
    C(\vec{r}-\vec{r}^\prime) = C_x(x-x')C_y(y-y') \\
\shortintertext{with}
    \label{eq:C:b}
    C_i(x) = \sinc^2\left(\frac{x}{\sigma_i}\right) \,, 
\end{gather}
\end{subequations}
where $i = x, y$ and we use $\sinc(x) = \sin(\pi x)/(\pi x)$.
Equivalently, the Fourier transform of the autocorrelation function yields the disorder power spectrum
\begin{subequations}
    \label{eq:C_x}
\begin{gather}
    \label{eq:C_x:a}
    \frac{\ensembleaverage{|V(\vec{k})-V_0|^2}}{L^2} = V_0^2 \sigma_x\sigma_y \tilde{C}_x(k_x) \tilde{C}_y(k_y) \\
\shortintertext{with}
    \label{eq:C_x:b}
    \tilde{C}_{i}(k) = \bigg( 1-\frac{|k \sigma_i|}{2\pi} \bigg) \theta\bigg(1 - \frac{|k \sigma_i|}{2\pi}\bigg) \,,
\end{gather}
\end{subequations}
where $V(\vec{k}) = \int \diff \vec{r} \, V(\vec{r}) \etothe{-i \vec{k} \cdot \vec{r}}$ is the Fourier transform of the disorder potential and $\theta(x)$ denotes the Heaviside step function.
The disorder correlation functions in Fourier and real space are illustrated in \cref{fig:speckle_disorder:a,fig:speckle_disorder:b}, respectively.

In addition to tuning the disorder strength $V_0$, it is also possible to create anisotropic disorder and even change the dimensionality of the disorder by choosing the correlation lengths $\sigma_i$ appropriately.
While allowing $\vec{k}$ modes up to a certain magnitude along both directions creates two-dimensional disorder that is on average isotropic ($\sigma_x = \sigma_y$), see \cref{fig:speckle_disorder:c,fig:speckle_disorder:e}, choosing only $\vec{k}$ modes along the $x$ direction creates a stripe-like disorder which we call one-dimensional disorder ($\sigma_y = L$), see \cref{fig:speckle_disorder:d,fig:speckle_disorder:f}.
In what follows, we will discuss both cases.

We note that the extension of our results from periodic to closed hard-wall boundary conditions, which are more practical for experiments, is straightforward. Changing the boundary conditions would produce different finite-size effects, but would not affect bulk properties.

\section{Theoretical Methods}
\label{sec:TM}

\begin{figure}[!htb]
    \includegraphics[width=\columnwidth]{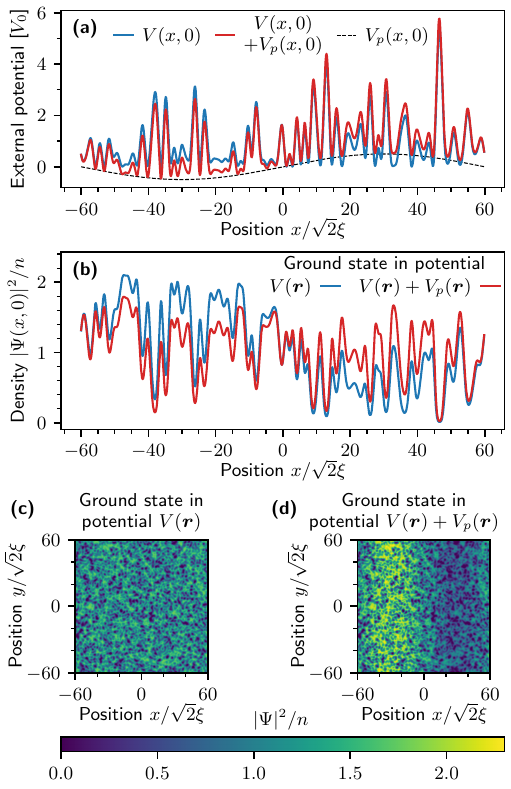}%
    \subfloat{\label{fig:density_wave:a}}%
    \subfloat{\label{fig:density_wave:b}}%
    \subfloat{\label{fig:density_wave:c}}%
    \subfloat{\label{fig:density_wave:d}}%
    \caption{\label{fig:density_wave}%
        Illustration of the linear-response protocol for probing sound propagation.
        The blue line in panel (a) is a cross section of the disorder potential~$V(\vec{r})$ at $y=0$.
        We solve the stationary Gross--Pitaevskii equation (GPE) to obtain the ground state in such a potential, yielding the density profile shown as a blue line in panel~(b) and as a color plot in panel~(c).
        We then calculate a new ground state with the same disorder, but adding also a sinusoidal probe potential~$V_p(\vec{r})$, shown as the black dashed line in panel~(a) for the wave vector $\vec{q} = \vec{\hat{e}_x} \mathinner{2 \pi / L}$.
        The perturbed potential $V(\vec{r}) + V_p(\vec{r})$ and the resulting ground-state density profile are depicted as red lines in panels (a) and (b), respectively, while panel~(d) shows the corresponding density as a color plot.
        This new ground state is used both as a tool to determine the density--density response function and as the initial configuration of a real-time evolution in the same disorder potential, after the sudden removal of the probe potential.   
        In this figure, $V_p(\vec{r})$ is large ($\lambda = 1/2$) for making the effects more visible, but the actual calculations are performed with weaker probe potentials ($\lambda \ll 1$).
        The disorder parameters are: $V_0 = gn$ and $\sigma_x = \sigma_y = 2 \sqrt{2} \xi$, where $gn$ is the bare chemical potential and $\xi$ is the healing length of the superfluid.%
    }
\end{figure}

The system of interest is a weakly interacting Bose--Einstein condensate (BEC) at zero temperature, described by the Gross--Pitaevskii energy functional
\begin{equation}
\label{eq:GPenergy}
    E = \int \diff\vec{r} \, \left\{ \Psi^*(\vec{r}) \left[ -\frac{\hbar^2 \nabla^2}{2m} + V(\vec{r}) \right] \Psi(\vec{r}) + \frac{g}{2} \abs*{\Psi(\vec{r})}^4 \right\} \,,
\end{equation}
where $\vec{r}=(x,y)$ is the position in two dimensions, $\Psi(\vec{r})$ is the condensate wave function, $V(\vec{r})$ is the disorder potential, and $g$ is the mean-field interaction strength.
In what follows, we denote the local density of the gas by $n(\vec{r}) = \abs{\Psi(\vec{r})}^2$, which is normalized to the total number of particles $N = \int \diff\vec{r} \, n(\vec{r})$.
The time evolution of the system is governed by the time-dependent GPE
\begin{equation}
    i \hbar \partial_t \Psi(\vec{r},t) = \left[-\frac{\hbar^2 \nabla^2}{2m} + V(\vec{r}) + g \abs*{\Psi (\vec{r},t)}^2 \right] \Psi(\vec{r},t) \,,
    \label{eq:GPE}
\end{equation}
which can be derived by applying the variational principle $i \hbar \partial_t \Psi = \delta E / \delta \Psi^*$ to the energy functional~\labelcref{eq:GPenergy}.

One can define the average density~$n=N/L^2$, the bare chemical potential~$gn$, and the healing length~$\xi = \hbar / \sqrt{2 m gn}$.
It is then convenient to use these quantities to work in natural units, where energies are expressed in units of $gn$, lengths in units of $\sqrt{2} \xi$, and time in units of $\hbar / gn$.

Flat two-dimensional traps can be realized by squeezing the gas into a transverse harmonic confinement of frequency~$\omega_z$. If $\omega_z$ is large enough, the motion along the third spatial dimension is frozen into the ground state of the harmonic potential~\cite{PhysRevLett.121.145301, Christodoulou2021}. The effective 2D interaction strength for such a system is $g = \tilde{g} \mathinner{\hbar^2 / m}$ with $\tilde{g} = \sqrt{8 \pi} a_s / a_z$, where $a_s$ is the three-dimensional $s$-wave scattering length and $a_z = \sqrt{\hbar / m \omega_z}$ is the oscillator length. In our calculations, we choose physical parameters close to the experiment in Ref.~\cite{Chauveau2023}, namely,
$\omega_z / 2 \pi = \SI{3.7}{\kilo\hertz}$ and 
$a_s = \SI{5.3}{\nano\meter}$ (for \rubidium atoms).
This yields~$\tilde{g} \approx \num{0.15}$.
Furthermore, we typically consider $N = \num{9.6e4}$ particles in a square box of length $L = \SI{40}{\micro\meter}$, corresponding to a mean particle density of $n = \SI{60}{\per\square\micro\meter}$.
The interaction strength is thus characterized by the dimensionless quantity $\tilde{g} N = 14400$, which, due to scale invariance in 2D, also determines the system size in natural length units, $L / \sqrt{2} \xi = \sqrt{\tilde{g} N} = 120$.

Given the above parameters, we will assume to work in the regime where there is a single condensate, such that the system is neither in a localized phase~\cite{Billy2008,Roati2008,White2020}, where there is no sound propagation, nor in a Bose glass phase~\cite{Riegler97,Fallan2007,Zuniga2015}, where the system is described by several incoherent condensates.
We note that in the conditions outlined above, the GPE provides an accurate description of the condensate wave function, identified as the eigenfunction with the largest eigenvalue of the one-body density matrix $\rho(\vec{r},\vec{r}^\prime)=\braket{0 | \hat{\psi}^\dagger(\vec{r}) \hat{\psi}(\vec{r}^\prime) |0}$, calculated in terms of the creation and annihilation field operators $\hat{\psi}^\dagger(\vec{r})$ and $\hat{\psi}(\vec{r})$ acting on the many-body ground state~$\ket{0}$. It is important to point out that, in the presence of broken translational symmetry, the occupation of the condensate wave function can be significantly different from the occupation of the $\vec{k}=0$ mode, which would coincide with the condensate in infinite uniform systems. 

Our primary interest is how the disorder will affect the speed of sound, the superfluid fraction, and the validity of hydrodynamic theory. As stated previously, the applicability of hydrodynamics implies a fundamental relation between the speed of sound, the superfluid fraction, and the inverse compressibility, \cref{cHD}. For this reason, we will also investigate how the compressibility changes with increasing disorder strength.

Our strategy is first to examine the equilibrium properties of the condensate, i.e., the compressibility and the superfluid fraction. These can readily be computed by considering the static response of the condensate to a long-wavelength density perturbation and a phase twist, respectively. These two quantities then give us access to the speed of sound via the hydrodynamic relation in \cref{cHD}. Secondly, we perform time-dependent simulations of the GPE after the sudden removal of a sinusoidal perturbation to extract the speed of sound and its damping.
The comparison between the equilibrium and dynamic simulations of the GPE allows us to further test the applicability of hydrodynamics. 

\subsection{Stationary Gross--Pitaevskii equation}
\label{sec:TM:stationary}

We obtain the equilibrium properties of the BEC in the presence of disorder by 
numerically solving the stationary GPE
\begin{equation}
    \label{eq:stationaryGPE}
    \mu \Psi(\vec{r}) = \left[ -\frac{\hbar^2 \nabla^2}{2m} + V(\vec{r}) + g \abs*{\Psi(\vec{r})}^2 \right] \Psi(\vec{r}) \,.
\end{equation}
This yields the ground-state condensate wave function~$\Psi(\vec{r})$ as well as the chemical potential~$\mu$.

\subsubsection{Compressibility}

To extract the inverse compressibility, we add a sinusoidal probe potential,
\begin{equation}
    V_p(\vec{r}) = \lambda gn \sin\left(\vec{q} \cdot \vec{r}\right) \,,
    \label{eq:V_probe}
\end{equation}
with strength $\abs{\lambda} \ll 1$ and wave vector $\vec{q}$, and obtain the ground state in the presence of both the disorder and the probe potential, as illustrated in \cref{fig:density_wave}.
If the magnitude of $\lambda$ is sufficiently small, this provides access to the static density--density response function $\chi(\vec{q})$, which, in the $q = |\vec{q}| \to 0$ limit,  is related to $\kappa$ by the compressibility sum rule~\cite{pitaevskii2016bose}:
\begin{equation}
\label{kappa}
\begin{split}
\quantumperturbed{\sin(\vec{q} \cdot \vec{r})} - \quantumequilibrium{\sin(\vec{q} \cdot \vec{r})} &= \lambda gn \int \diff \omega \, \frac{S(\vec{q},\omega)}{\omega} \\
&= \lambda gn \frac{\chi(\vec{q})}{2}
\overset{\vec{q} \to 0}{=} \frac{\lambda}{2} gn^2 \kappa \,,
\end{split}
\end{equation}
where $S(\vec{q},\omega)$ is the dynamic structure factor.
Here and in what follows, $\quantumaverage{O} = N^{-1} \int \mathrm{d}\vec{r} \, \Psi^*(\vec{r}) O \Psi(\vec{r})$ denotes the quantum average of an observable~$O$, while we use the subscripts $\quantumequilibrium{}$ and $\quantumperturbed{}$ to emphasize expectation values with respect to the ground state in absence and presence of the sinusoidal perturbation, respectively.
In the following, we will always identify $\chi(\vec{q})$ with $n \kappa$, assuming that $q = 2 \pi / L$ is sufficiently small.

\subsubsection{Superfluid fraction}

The superfluid fraction can readily be extracted using the phase-twist method~\cite{PhysRevA.8.1111,PhysRevA.99.041601}.
To this end, we perform a gauge transformation on the condensate wave function, $\Psi(\vec{r},t) \rightarrow \etothe{i m \vec{v}_s \cdot \vec{r}/\hbar} \Psi(\vec{r},t)$, with  the superfluid velocity~$\vec{v}_s$. The superfluid fraction~$f_s$ in the direction of the superfluid flow can then be calculated by examining how the energy~\labelcref{eq:GPenergy} changes in response to the velocity field in the limit $v_s \to 0$,
\begin{equation}
\label{eq:phasetwistenergy}
    E = E_0 + N f_s \frac{1}{2}m v_s^2 \,,
\end{equation}
where $E_0$ is the energy in the absence of the phase twist.

As a consistency check, we also consider the validity of Leggett's bounds~\cite{Leggett1970, Leggett1998}.
Leggett showed that the superfluid fraction satisfies the inequality
\begin{equation}
    f_{s}^- \le f_{s} \le f_{s}^+ \,,
\end{equation} 
where the first inequality ($f_{s}^- \le f_{s}$) is applicable only to a dilute Bose gas described by the GPE. The bounds on the superfluid fraction, $f_{s}^{\pm}$, are defined in terms of the density of the atomic gas. For the case of our two-dimensional condensate, we have
\begin{subequations}
\label{eq:superfluid_bounds}
\begin{align}
\label{eq:superfluid_bounds:lower}
f_s^- &= \frac{1}{n} \int \frac{\diff r_{\perp}}{L} \left[\int \frac{\diff r_{\|}}{L} \frac{1}{n(\vec{r})} \right]^{-1} \,, \\
\label{eq:superfluid_bounds:upper}
f_s^+ &= \frac{1}{n} \left[ \int \frac{\diff r_{\|}}{L} \frac{1}{\int \frac{\diff r_{\perp}}{L}n(\vec{ r})} \right]^{-1} \,.
\end{align}
\end{subequations}
Here, $r_{\| (\perp)}$ are the directions parallel (perpendicular) to the superfluid flow, and $n$ (without the position index) is the average density $n= N/L^2$. In general, these two bounds differ from one another when the density varies in both directions, which can happen in the presence of two-dimensional disorder potentials. If the disorder potential is one-dimensional, i.e., only a function of one direction, the two Leggett bounds collapse onto one another, and \cref{eq:superfluid_bounds} provides a second efficient method for calculating the superfluid fraction in systems described by the GPE.

\subsection{Time-dependent Gross--Pitaevskii equation}
\label{sec:TM:dynamical}

The equilibrium simulations described in the previous section allow one to extract the speed of sound, thanks to \cref{cHD}. However, it is not known a priori how well the hydrodynamic formula reproduces the dynamics in the regime of strong disorder.
For this reason, we also perform time-dependent simulations of the GPE to extract the speed of sound and its damping directly.

We excite the sound mode in a manner similar to the study of the compressibility.
First, the ground state is determined in the presence of both the disorder and the probe potential in \cref{eq:V_probe}, as illustrated in \cref{fig:density_wave}.
At time $t=0$, the probe potential is removed and the dynamics is studied by numerically evaluating the time-dependent GPE.

The relevant observable is $\ensembleaverage{\quantumaverage*{\sin(\vec{q} \cdot \vec{r})}}(t)$, the disorder-averaged expectation value of $\sin(\vec{q} \cdot \vec{r})$.
We expect the dynamics of this quantity to be readily fit by damped oscillations of the form
\begin{equation}
    \label{eq:sound_fit}
    \ensembleaverage{\quantumaverage*{\sin(\vec{q} \cdot \vec{r})}}(t) = \frac{1}{2} \lambda gn \chi(\vec{q}) \etothe{- \gamma_{\vec{q}} t} \cos(\omega_{\vec{q}} t) + \ensembleaverage{\quantumequilibrium*{\sin(\vec{q} \cdot \vec{r})}} \,,
\end{equation}
where $\omega_{\vec{q}} = c q$ is the dispersion relation of the sound mode and $\gamma_{\vec{q}}$ is the associated damping rate.
We will also assess the strength of the damping via the quality factor $Q_{\vec{q}} = \omega_{\vec{q}} / 2 \gamma_{\vec{q}}$ of the oscillations.
Notably, the amplitude of the oscillations is set by the density response function $\chi(\vec{q})$, or equivalently, by the compressibility, see \cref{kappa}.
Statistical error bars of the fit parameters are estimated using jackknife resampling~\cite{Young2015}.

\subsection{Perturbation theory}
\label{sec:PT}
In the limit of weak disorder, where the disorder strength is much less than the bare chemical potential, $V_0 \ll gn$, perturbation theory is applicable. Numerous studies have obtained analytical results for thermodynamic and transport properties of the gas~\cite{Huang1992,Giorgini1994, Falco2007,Astrakharchik2013,Gaul2014,Cappellaro2019}.
As discussed in \cref{app:PT}, it is possible to expand the condensate wave function in powers of $V_0 / gn$, and to obtain various physical observables.
First, consider the chemical potential,
\begin{equation}
    \frac{\mu}{gn} = 1 + \frac{\ensembleaverage{V(\vec{k}=0)}}{gn L^2} - \frac{1}{gn L^2}\sum_{\vec{k}} \frac{\ensembleaverage{\left|V(\vec{k})\right|^2}}{L^2} \frac{\epsilon_{\vec{k}}^3}{E_{\vec{k}}^4} \,.
    \label{eq:chemical_potential}
\end{equation}
We also define the single-particle energy $\epsilon_{\vec{k}} = \hbar^2 k^2/2m$ and the Bogoliubov quasiparticle energy $E_{\vec{k}} = \sqrt{\epsilon_{\vec{k}} (\epsilon_{\vec{k}} + 2gn)}$.
The change in the chemical potential contains two pieces: a trivial shift of the chemical potential due to the mean disorder strength, $\ensembleaverage{V(\vec{k}=0)} / L^2 = V_0$, and a quadratic shift which is negative. From \cref{eq:chemical_potential}, it is straightforward to calculate the inverse compressibility $\kappa^{-1}=n^2\partial \mu/\partial n$, yielding
\begin{equation}
\label{eq:compressibility}
    \frac{\kappa^{-1}}{gn^2} = 1 + \frac{4}{L^2}\sum_{\vec{k}} \frac{\ensembleaverage{\left|V(\vec{k})\right|^2}}{L^2}\frac{\epsilon_{\vec{k}}^4}{E_{\vec{k}}^6} \,.
\end{equation}
The inverse compressibility is quadratic in the disorder and is strictly positive, i.e., the gas becomes more incompressible with increasing disorder strength.

As discussed previously, we extract the superfluid fraction by performing the phase-twist method. In the GPE, this shifts the chemical potential by $m v_s^2 / 2$ and the momentum by $m v_s$. We can again solve the GPE perturbatively in $V_0/gn$ and evaluate the particle current $J_i = N f_s^{i,j} v_s^j$. The result of the perturbative calculation is:
\begin{equation}
    f_s^{i,j} = \delta_{i,j} - \frac{4}{L^2}\sum_{\vec{k}} \frac{\ensembleaverage{\left|V(\vec{k})\right|^2}}{L^2} \frac{\epsilon_{\vec{k}}^2}{E_{\vec{k}}^4} \hat{k}_i \hat{k}_j \,, \label{eq:superfluid_density}
\end{equation}
where $\hat{k}_i$ is the $i$-th component of the unit vector~$\vec{\hat{k}} = \vec{k} / k$.
This result appeared before in the literature, see, e.g., Ref.~\cite{Astrakharchik2013}. In general, the superfluid fraction is a tensor $f_s^{i,j}$ with $i,j = x,y$. In the absence of disorder, the superfluid fraction is isotropic and is unity due to Galilean invariance. However, depending on whether the disorder is isotropic or anisotropic, the superfluid fraction can be isotropic or anisotropic, respectively.

We finally calculate the speed of sound and its damping rate. We calculate the sound mode by invoking hydrodynamics. As described in \cref{app:PT}, linearizing the GPE for small density and phase fluctuations yields a sound mode of the form~\labelcref{cHD}, but now with the speed of sound promoted to a tensor. The result for the speed-of-sound tensor~$c_{i, j}$ according to perturbation theory is:
\begin{equation}
\label{eq:speed_of_sound}
\begin{split}
    \frac{c^2_{i,j}}{c_0^2} &= \delta_{i,j}\left(1 + \frac{4}{L^2}\sum_{\vec{k}} \frac{\ensembleaverage{\left|V(\vec{k})\right|^2}}{L^2} \frac{\epsilon_{\vec{k}}^4}{E_{\vec{k}}^6} \right) \\
    &\hphantom{={}} - \frac{4}{L^2}\sum_{\vec{k}}\ \frac{\ensembleaverage{\left|V(\vec{k})\right|^2}}{L^2} \frac{\epsilon_{\vec{k}}^2}{E_{\vec{k}}^4} \hat{k}_i \hat{k}_j \,,
\end{split}
\end{equation}
where $c_0 = \sqrt{gn / m}$ is the speed of sound in the absence of disorder.
This expression is just a combination of \cref{eq:compressibility,eq:superfluid_density}.

The presence of a sound mode implies that the dynamic structure factor at small momentum is given by~\cite{Giorgini1994}:
\begin{equation}
    S(\vec{q},\omega) \approx \frac{\vec{q} \cdot \vec{n}_s \cdot \vec{q}}{\pi}\frac{2 c_0 q \gamma_{\vec{q}}}{(\omega^2 - \vec{q} \cdot \vec{c}^2 \cdot \vec{q})^2+ 4(c_0 q)^2 \gamma_{\vec{q}}^2} \,,
\end{equation}
where $\gamma_{\vec{q}}$ is the damping of the sound mode with momentum $\vec{q}$ and $\vec{q} \cdot \vec{n}_s \cdot \vec{q} = \sum_{i,j} q_i q_j n_s^{i,j}$ denotes the tensor product (similarly for $\vec{q} \cdot \vec{c}^2 \cdot \vec{q}$).
In \cref{app:damping}, we perform a perturbative calculation for the damping, which yields the result~\cite{Giorgini1994}
\begin{equation}
\label{eq:damping_rate}
\frac{\gamma_{\vec{q}}}{c_0 q} = \frac{1}{8} \frac{q^2}{(gn)^2} \int \frac{\diff \theta_{\vec{k}}}{2\pi} \frac{\ensembleaverage{\abs[\big]{V \big( q \vec{\hat{k}} \big)}^2}}{L^2} \big( \vec{\hat{q}} \cdot \vec{\hat{k}} \big)^2 \,,
\end{equation}
where $\vec{\hat{q}} = \vec{q} / q$ and $\theta_{\vec{k}}$ is the angle of the unit vector $\vec{\hat{k}}$ with respect to the $x$~axis.


\section{Results: Two-dimensional Disorder}
\label{sec:2D_disorder}

\begin{figure*}[htb]
	\includegraphics[width=\linewidth]{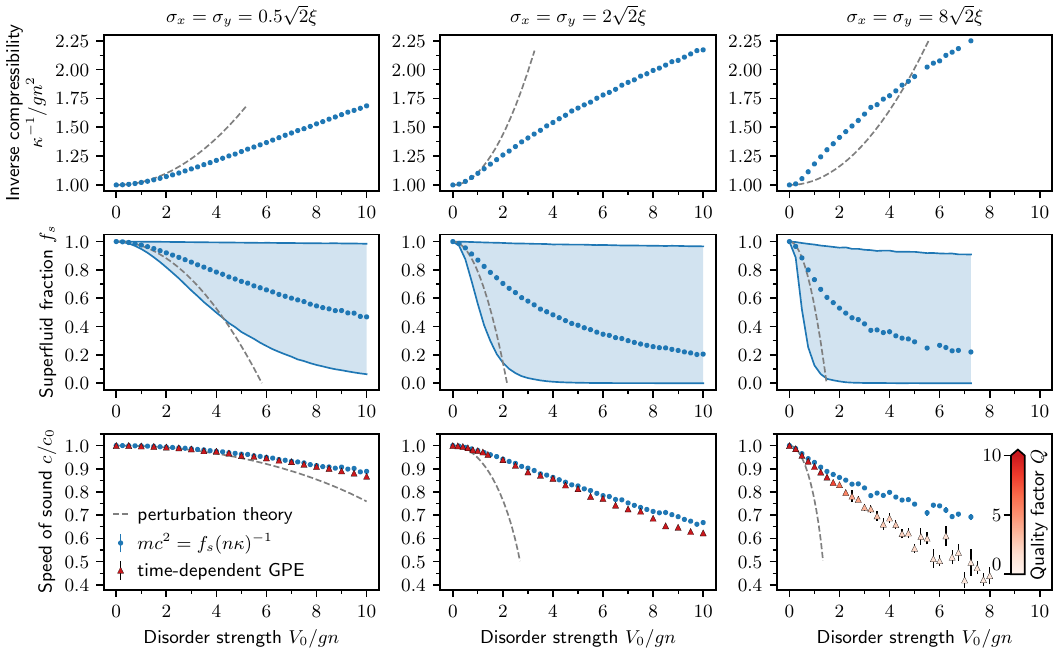}%
	\caption{\label{fig:superfluidity_2d}%
        Equilibrium and transport properties of a two-dimensional Bose--Einstein condensate subject to isotropic 2D speckle disorder as a function of the disorder strength~$V_0$.
        Each column corresponds to a different value of the disorder correlation length $\sigma_x = \sigma_y$.
        The blue dots in the top row represent the inverse compressibility~$\kappa^{-1}$, obtained from the static linear response according to \cref{kappa}, while in the middle row they represent the superfluid fraction~$f_s$, computed using the phase-twist method. The solid lines correspond to Leggett's bounds, which become loose already for moderately strong disorder due to the non-separability of the density profile. In the bottom row, the blue dots represent the speed of sound obtained from the hydrodynamic relation~\labelcref{cHD}, while the triangles are the values extracted from time-dependent GPE simulations.
        The error bars show the standard error of the mean. The red color scale used for the triangles encodes the quality factor~$Q$ of the damped oscillations after exciting a phonon with wave number $q = 2 \pi / L$.
        The gray dashed lines show the predictions from perturbation theory for the compressibility, superfluid fraction, and speed of sound, as given by \cref{eq:compressibility,eq:superfluid_density,eq:speed_of_sound}, respectively.
        The physical quantities extracted from stationary (time-dependent) GPE simulations have been averaged over $16, 32, 64$ ($16, 16, 64$) realizations for the three values of the correlation length $\sigma_x = \sigma_y = (0.5, 2, 8) \sqrt{2} \xi$, respectively.%
    }
\end{figure*}

\begin{figure}[!htb]
	\includegraphics[width=\columnwidth]{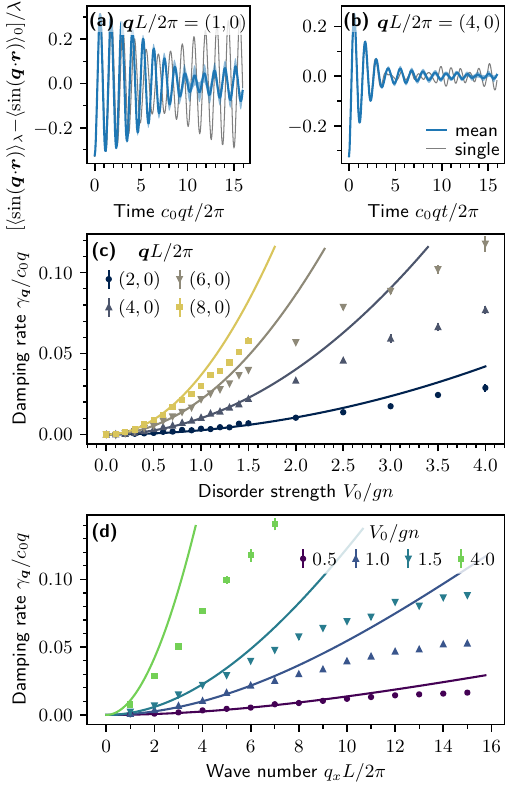}%
    \subfloat{\label{fig:damping_2d:a}}%
    \subfloat{\label{fig:damping_2d:b}}%
    \subfloat{\label{fig:damping_2d:c}}%
    \subfloat{\label{fig:damping_2d:d}}%
	\caption{\label{fig:damping_2d}%
        Damping of sound waves for isotropic 2D speckle disorder.
        (a),(b)~Time trace of the observable $\braket{\sin(\vec{q} \cdot \vec{r})}$ after suddenly releasing the probe potential for two different wave vectors $\vec{q} = (q_x, q_y)$ with disorder strength $V_0 / gn = 4$ and correlation length $\sigma_x = \sigma_y = 2 \sqrt{2} \xi$. The blue curve shows the ensemble average over $16$ realizations with the shaded area corresponding to the standard error of the mean, while the gray line depicts a typical single realization.
        If the longest wavelength commensurate with the box is excited~(a), a single realization is characterized by sizable revivals, which vanish in the ensemble average due to dephasing. Revivals are instead suppressed for phonons at shorter wavelengths~(b) even in single trajectories.
        (c)~Damping rate~$\gamma_{\vec{q}}$ as a function of the disorder strength for several wave vectors, extracted by fitting an exponentially damped sinusoidal oscillation, \cref{eq:sound_fit}, to the ensemble-averaged signal.
        (d)~$q_x$-dependence of the damping rate for several disorder strengths.
        The damping rate predicted by perturbation theory according to \cref{eq:damping_rate} (solid lines) reproduces the numerical results well for moderately strong disorder and small wave vectors.%
    }
\end{figure}

We start our analysis by investigating the case of isotropic 2D disorder, i.e., disorder applied in both the $x$ and $y$ directions with the same correlation lengths, $\sigma_x = \sigma_y$. 
Examples of such potentials are shown in \cref{fig:speckle_disorder:c,fig:speckle_disorder:e}, while a typical ground-state density profile is depicted in \cref{fig:density_wave:c}.
To probe equilibrium properties and sound propagation in this setup, we employ the linear-response protocol illustrated in \cref{fig:density_wave} and described in \cref{sec:TM:stationary,sec:TM:dynamical}.
Note that in \cref{fig:density_wave:b,fig:density_wave:d}, the strength of the probe potential for exciting sound waves is exaggerated for a better visual impression; the results presented below have been obtained for much weaker perturbations ($\lambda \lesssim \num{0.01}$) well inside the linear-response regime.

\Cref{fig:superfluidity_2d} summarizes our results for the compressibility~$\kappa$, the superfluid fraction~$f_s$, and the speed of sound~$c$ as a function of the disorder strength~$V_0$ for three values of the correlation length $\sigma_x = \sigma_y$.
The numerical results are compared to the predictions of perturbation theory for the respective quantities, see \cref{eq:compressibility,eq:superfluid_density,eq:speed_of_sound}.

\subsection{Compressibility}
The inverse compressibility reported in \cref{fig:superfluidity_2d} has been obtained from the static linear response according to \cref{kappa} for $\vec{q} = \vec{\hat{e}}_x \mathinner{2 \pi / L}$.
In practice, we extract  $\kappa^{-1}$ from the stationary GPE~\labelcref{eq:stationaryGPE} for fixed disorder strength~$V_0$ and correlation length $\sigma_x = \sigma_y$, and for several realizations of the disorder potential.
The blue points correspond to the ensemble averages (statistical error bars showing the standard error of the mean are often smaller than or equal to the marker size and therefore not visible).
We have also checked that the values of $\kappa^{-1}$ obtained in this way agree with the thermodynamic expression $\kappa^{-1} = n^2 \partial \mu / \partial n$ within the numerical uncertainties. 
Notably, the \emph{inverse} compressibility always increases with the disorder strength, i.e., the compressibility decreases. 
Perturbation theory correctly reproduces this behavior for weak disorder, but the range of validity shrinks for larger values of the correlation length.

\subsection{Superfluid fraction}
The superfluid fraction in \cref{fig:superfluidity_2d} has been calculated using the phase-twist method, see \cref{eq:phasetwistenergy}, and is shown together with Leggett’s upper and lower bounds, computed from the equilibrium density profile according to \cref{eq:superfluid_bounds:upper,eq:superfluid_bounds:lower}.
We find that the two bounds differ significantly already for moderately strong disorder, which is related to the non-separability of the equilibrium density profile caused by the 2D nature of the disorder.
Thus, for our setup, Leggett's bounds are too imprecise for estimating the superfluid fraction, requiring alternative approaches, e.g., based on the hydrodynamic relation~\labelcref{cHD}, as discussed below.
Note that, with respect to the recent results of Ref.~\cite{perezcruz2024}, where the role of disorder has been studied in a similar geometry, the two Leggett bounds differ more significantly in our case due to the larger value of the interaction strength~$\tilde{g} N$ used in our simulations.

\subsection{Speed of sound}
In the lower panels of \Cref{fig:superfluidity_2d}, we compare the speed of sound extracted from the compressibility and the superfluid fraction by virtue of the hydrodynamic relation~\labelcref{cHD} with the value obtained from time-dependent GPE simulations by using the fitting function~\labelcref{eq:sound_fit}.
It can be seen that the hydrodynamic predictions agree well with the time-dependent GPE simulations across a wide range of disorder strengths and correlation lengths.
This suggests that, in this regime, the hydrodynamic relation~\labelcref{cHD} can be used as a tool to measure the superfluid fraction, especially in situations where Leggett's bounds are instead too loose.
We remark that the required linear-response measurement of the compressibility and the speed of sound is feasible with modern experimental setups similar to the one in Ref.~\cite{Chauveau2023}.

\subsection{Damping of sound waves}
By fitting the disorder-averaged expectation value $\ensembleaverage{\quantumaverage{\sin({\vec{q} \cdot \vec{r}})}}(t)$ obtained from time-dependent GPE simulations, see \cref{eq:sound_fit}, we can also extract the damping rate~$\gamma_{\vec{q}}$ of sound waves as well as and the corresponding quality factor $Q_{\vec{q}} = \omega_{\vec{q}} / 2 \gamma_{\vec{q}}$.
In the lower panels of \cref{fig:superfluidity_2d}, the quality factor for $\vec{q} = \vec{\hat{e}}_x \mathinner{2 \pi / L}$ is represented by the color scale of the red triangles.
One can see that the parameter range where the damping is small ($Q \gg 1$) is the same where the hydrodynamic prediction for the speed of sound is accurate.
If the disorder strength and the correlation length become too large, sound waves are strongly damped, up to the point where the sound mode is completely destroyed, as indicated by the quality factor dropping to a value on the order of one.

In \cref{fig:damping_2d}, we investigate the damping of sound waves in further detail.
The numerical results for the damping rate as a function of the disorder strength~$V_0$ and the phonon wave vector $\vec{q} = q_x \vec{\hat{e}}_x$, see \cref{fig:damping_2d:c,fig:damping_2d:d}, respectively, are well reproduced by the perturbative prediction in \cref{eq:damping_rate} for sufficiently weak disorder and small phonon wave vectors. 
It is worth stressing that here, damping is not due to incoherent collisions (which are not included in GPE), but is the consequence of the coherent coupling of Bogoliubov excitations induced by the presence of disorder.
In particular, the disorder potential couples sound modes that are close in energy, over which the energy of the initial collective excitation spreads.
A consequence of this is that, in the case of small-$q_x$ phonons with wavelength comparable to the system size, performing the average over several disorder realizations is crucial in order to obtain a sufficiently accurate estimate of $\gamma_{\vec{q}}$.
In fact, the discreteness of the excitation spectrum at low momenta favors the coupling among few low-lying collective modes, yielding sizable revivals of the initial oscillations; these revivals vanish after averaging due to dephasing, as shown in \cref{fig:damping_2d:a}.
Phonons at shorter wavelengths exhibit damped oscillations even in single trajectories, as in \cref{fig:damping_2d:b}, with suppressed revivals due to the larger number of modes with similar energy coupled by the disorder potential.
A more detailed discussion of these finite-size effects can be found in \cref{app:finite_size}.

\subsection{Validity of hydrodynamics}
As a final remark, we notice that the applicability of the hydrodynamic relation~\labelcref{cHD} was not guaranteed \textit{a priori} since hydrodynamics is expected to hold only if the wavelength of the phonon, equal to the box size $L$ in our simulation, is much larger than the wavelengths characterizing the external perturbation.
In our setting, the speckle potential has Fourier components at all wave numbers up to the inverse correlation length, whose power is strongest at wave numbers comparable to that of the phonon ($2 \pi / L$), see \cref{fig:speckle_disorder:a}.
These low-$q$ components of the disorder power spectrum, $\ensembleaverage{\abs{V(q \vec{\hat{k}})}^2} / L^2$, constitute the leading contribution to the damping rate according to \cref{eq:damping_rate}.
Since $\ensembleaverage{\abs{V(\vec{k})}^2} / L^2 \propto V_0^2 \sigma_x \sigma_y$, see \cref{eq:C_x}, they become impactful for large values of the disorder strength and correlation length, which eventually causes the hydrodynamic description to break down.
As we will discuss next, for stripe-like 1D disorder, where $\sigma_y = L$, this breakdown occurs at lower values of $V_0$ compared to isotropic 2D disorder.
The wider validity regime of hydrodynamics in the latter case can be explained by the 2D nature of phase space, which leads to a distribution of the spectral power over a larger set of modes and thus reduces the effects of small-wave-vector components in the disorder direction.

\section{Results: One-dimensional Disorder}
\label{sec:1D_disorder}

\begin{figure*}[htb]
	\includegraphics[width=\linewidth]{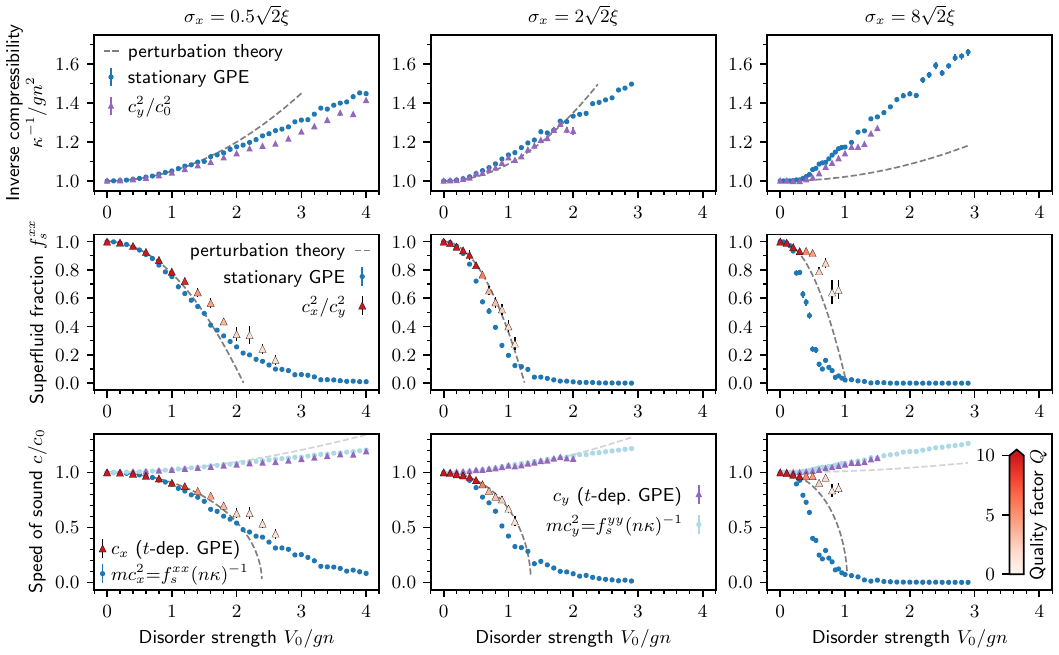}%
	\caption{\label{fig:superfluidity_1d}%
        Same as \cref{fig:superfluidity_2d}, but for stripe-like 1D speckle disorder.
        Since the system is fully superfluid in the $y$~direction ($f_s^{yy} = 1$), according to the hydrodynamic relation~\labelcref{cHD}, the speed of sound along $y$ directly yields the inverse compressibility, $(n \kappa)^{-1} = m c_y^2$ (purple triangles, upper row), while the ratio of the two sound velocities corresponds to the superfluid fraction in the $x$~direction, $f_s^{xx} = c_x^2 / c_y^2$ (reddish triangles, middle row).
        The color code shows the quality factor~$Q$ of the ensemble-averaged damped oscillations after exciting a phonon with wave vector $\vec{q} = \vec{\hat{e}_x} \mathinner{2 \pi / L}$.
        The hydrodynamic description is valid for weak disorder, but fails if the strength and the correlation length of the disorder potential become too large.
        The physical quantities extracted from stationary (time-dependent) GPE simulations have been averaged over $100, 100, 200$ ($50, 100$, $200$) realizations for the three values of the correlation length $\sigma_x = (0.5, 2, 8) \sqrt{2} \xi$, respectively.%
    }
\end{figure*}

\begin{figure}[!htb]
	\includegraphics[width=\columnwidth]{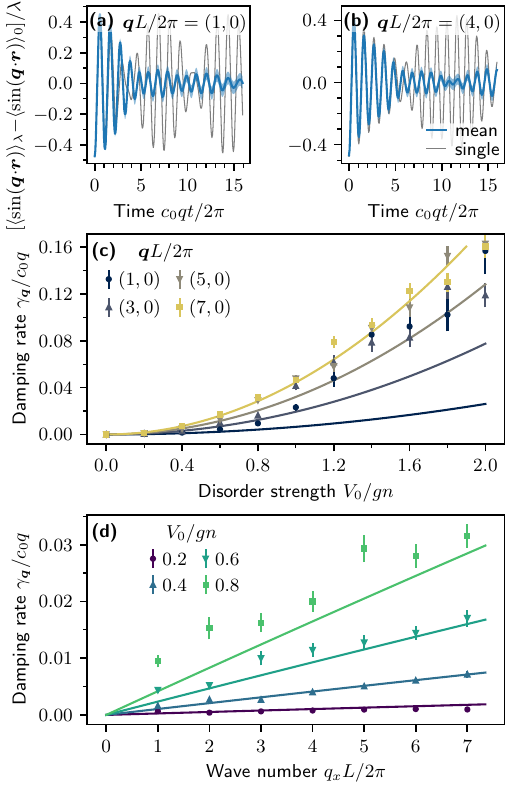}%
    \subfloat{\label{fig:damping_1d:a}}%
    \subfloat{\label{fig:damping_1d:b}}%
    \subfloat{\label{fig:damping_1d:c}}%
    \subfloat{\label{fig:damping_1d:d}}%
	\caption{\label{fig:damping_1d}%
        Same as \cref{fig:damping_2d}, but for stripe-like 1D speckle disorder with $\sigma_x = \num{0.5} \, \sqrt{2} \xi$.
        Unlike for isotropic 2D disorder, single realizations, shown for $V_0 / gn = 1$ as gray lines in panels $(a)$ and $(b)$, exhibit sizable revivals also at larger phonon wave numbers~$q_x$
        The ensemble averages (blue lines), computed over $50$ realizations, exhibit suppressed revivals due to dephasing.
        (c),(d)~The damping rates~$\gamma_{\vec{q}}$ extracted from the ensemble-averaged oscillations agree with the predictions from perturbation theory in \cref{eq:damping_rate} for sufficiently small disorder strengths~$V_0$ and phonon wave numbers~$q_x$.%
    }
\end{figure}

In this section, we discuss our numerical results for stripe-like 1D disorder.
In this case, the disorder potential only depends on the $x$~coordinate and has the form of disordered stripes, as illustrated in \cref{fig:speckle_disorder:d,fig:speckle_disorder:f}.
As a consequence, the ground state density profile is fully separable and constant along the stripes in the $y$ direction.
Due to the separability of the density, Leggett's upper and lower bounds in \cref{eq:superfluid_bounds} coincide~\cite{Chauveau2023}, and we have checked that the resulting value of the superfluid fraction agrees with the one obtained from the phase-twist method.
The following numerical results based on the stationary GPE use the superfluid fraction computed from the Leggett integral of the ground-state density profile.
As expected from the reduced number of random modes, 1D disorder requires averaging over a higher number of realizations in order to obtain results with similar statistical accuracy as in the 2D case.

\subsection{Compressibility, superfluid fraction, and speed of sound}

In \cref{fig:superfluidity_1d}, we report the analogous numerical results for the compressibility, superfluid fraction, and speed of sound as in \cref{fig:superfluidity_2d}, but for stripe-like 1D disorder.
As a general observation, we find that 1D disorder has more drastic effects on the superfluidity of the system as compared to 2D disorder.
This is intuitive as stripe-like potential barriers span the entire system along $y$, such that there is no way for the superfluid flow to circumvent obstacles as in the 2D case. 
For example, for $\sigma_x = 2 \sqrt{2} \xi$, in order to observe a $50\%$ reduction of the superfluid fraction with 1D disorder, a disorder strength on the order of the bare chemical potential ($V_0 \approx gn$) is enough, while with 2D disorder, one needs a four times larger strength.

The bottom row of \cref{fig:superfluidity_1d} shows the sound velocities $c_x$ and $c_y$, extracted from the response to suddenly releasing the probe potential with wave vector $\vec{q} = \vec{\hat{e}}_x \mathinner{2 \pi / L}$ and $\vec{q} = \vec{\hat{e}}_y \mathinner{2 \pi / L}$, respectively.
Due to the anisotropy of the 1D disorder potential, the sound velocities longitudinal and transversal to the disorder direction are different on average: while $c_x$ decreases with increasing disorder strength, $c_y$ is slightly enhanced.
The increase of the transversal sound velocity $c_y$ can be understood from the hydrodynamic relation~\labelcref{cHD}: since the density is constant along $y$, the system is fully superfluid in this direction, $f_s^{yy} = 1$, such that the increase in the speed of sound reflects the reduction of the compressibility, $m c_y^2 = (n \kappa)^{-1}$.

\subsection{Damping of sound waves}

In \cref{fig:damping_1d}, we analyze the damping of sound waves for 1D disorder in a similar fashion as for 2D disorder in \cref{fig:damping_2d}.
Unlike in the latter case, 1D disorder is characterized by sizable revivals in the sound-wave oscillations of single realizations even at larger phonon wave vectors, cf.~\cref{fig:damping_1d:a,fig:damping_1d:b}.
This behavior can be explained by the fact that the 1D disorder potential has momentum components in only one direction and therefore couples fewer collective modes as compared to 2D disorder.
Consequently, finite-size effects caused by the discreteness of the spectrum are more pronounced in the 1D disorder scenario (see \cref{app:finite_size} for more details).
Nonetheless, after averaging over many disorder realizations, the revivals in the sound-wave oscillations vanish due to dephasing and a damping rate can be extracted by fitting \cref{eq:sound_fit}, as before.
In \cref{fig:damping_1d:c,fig:damping_1d:d}, it can be seen that the perturbative prediction in \cref{eq:damping_rate} correctly reproduces the damping rate extracted this way for sufficiently weak disorder and small phonon wave numbers.%

\subsection{Validity of hydrodynamics}

One particular appeal of the 1D disorder scenario is the prospect of exploiting the hydrodynamic relation~\labelcref{cHD} in order to extract the superfluid fraction from the ratio of the squared sound velocities, $f_s^{xx} = c_x^2 / c_y^2$.
This method has been used in Ref.~\cite{Chauveau2023} for the measurement of the superfluid fraction in the presence of a 1D optical-lattice potential with a period much smaller than the box size.
While this technique is still applicable also for 1D speckle disorder, its accuracy is limited to moderate disorder strengths and small correlation lengths (see central row of \cref{fig:superfluidity_1d}) due to the discrepancies in the speed of sound described below.

In the top and bottom rows of \cref{fig:superfluidity_1d}, it can be seen that the hydrodynamic description is accurate in the $y$ direction across the investigated parameter regime.
Conversely, in the $x$ direction, the validity of hydrodynamics is more subtle: if the disorder is too strong, mode coupling effects become important and the speed of sound extracted from the time-dependent simulation deviates significantly from the value obtained using the hydrodynamic relation~\labelcref{cHD}.
As mentioned in \cref{sec:2D_disorder}, we attribute  the stronger effects of disorder in 1D to the enhanced spectral power of the disorder potential at small wave vectors~$\vec{k}$ comparable to the wave vector $\vec{q}$ of the phonon.
We have checked that by removing such long-wavelength components with a proper filter procedure, the consistency with the hydrodynamic description improves, see \cref{app:infrared_cutoff}.

\section{Conclusion}
\label{sec:conclusions}

In this paper, we have discussed the influence of disorder on the superfluid and dynamical properties of an interacting Bose--Einstein-condensed gas.
Our main motivation was to understand whether hydrodynamic theory of superfluids, yielding at zero temperature the fundamental relationship~\labelcref{cHD} between the sound velocity, superfluid density, and compressibility, provides a reliable description also in the presence of disorder.
The validity of hydrodynamic theory has been investigated by comparison with numerical simulations based on time-dependent Gross--Pitaevskii theory.
In what follows, we summarize the major results of our analysis for a two-dimensional BEC at zero temperature subject to speckle disorder.

Although disorder makes the system less compressible, it can result in a significant reduction of the sound velocity and a concomitant decline of the superfluid fraction, causing an increase of the inertia of the superfluid motion.
In addition, the sensitivity of the system to disorder depends crucially on the dimensionality of the disorder potential.
In particular, if disorder is applied in only one direction (1D disorder), the speed of sound acquires an anisotropic character and the effects on the superfluid fraction (and consequently on the sound velocity) become important also at moderate strengths of the disorder potential.
Moreover, the superfluid hydrodynamic relation~\labelcref{cHD} turns out to be accurate across a wide range of the explored parameters, also beyond the applicability of perturbation theory.
It consequently allows for a precise experimental estimate of the superfluid fraction, especially in the presence of 2D isotropic disorder, where alternative approaches, e.g., based on the use of Leggett's bounds, fail.
However, the hydrodynamic description becomes inaccurate if the disorder is very strong, and this breakdown is more drastic in 1D and for larger correlation lengths of the disorder potential.
Finally, disorder introduces collisionless damping of sound waves via mode coupling.
We have also identified relevant finite-size effects leading to revivals of oscillations in individual realizations, especially for 1D disorder and for phonons with wavelength on the order of the system size.

Further questions, which remain to be explored in future works, concern the study of other types of disorder potentials, e.g. random piecewise constant potentials, which can be generated experimentally using digital micromirror devices~\cite{Dalibard2024}.
Another important issue concerns the effect of disorder on the quantum fluctuations of the condensate and the corresponding consequences on the propagation of sound.

\appendix

\begin{acknowledgments}
We thank G.\ Astrakharchik, J.\ Dalibard, J.\ Beugnon, S.\ Nascimb\`ene, F.\ Rabec, P.\ Massignan,  D.\ P\'erez-Cruz, and J.\ Spielman for useful discussions.
KTG would like to thank the Coll\`ege de France and the team of J.\ Dalibard at the Laboratoire Kastler Brossel for the kind hospitality.
AB would like to thank the Institut Henri Poincaré (UAR 839 CNRS-Sorbonne Université) and the LabEx CARMIN (ANR-10-LABX-59-01) for their support.
This project has received funding from the European Research Council (ERC) under the European Union’s Horizon 2020 research and innovation programme (Grant Agreement No.\ 804305) and by the European Union --- NextGeneration EU, within PRIN 2022, PNRR M4C2, Project TANQU 2022FLSPAJ [CUP B53D23005130006].
This work has benefited from Q@TN, the joint lab between the University of Trento, FBK---Fondazione Bruno Kessler, INFN---National Institute for Nuclear Physics, and CNR---National Research Council.
We further acknowledge support by Provincia autonoma di Trento.
\end{acknowledgments}

\section{Sampling speckle disorder}
\label{app:disorder_protocol}

In this appendix, we discuss a discrete version of the procedure described in Ref.~\cite{Pilati2010} for sampling speckle potentials.
The main difference between our approach and Ref.~\cite{Pilati2010} is that here we sample only a discrete set of Fourier modes whose wave vectors are commensurate with the size of the box.
This way, the speckle potential satisfies periodic boundary conditions by construction.

We start by defining the amplitude field
\begin{equation}
\label{eq:speckle_amplitude}
    A(\vec{r}) = \sum_{\vec{k}} a_{\vec{k}} \etothe{i \vec{k} \cdot (\vec{r} - \vec{r}_0)}
\end{equation}
on a $d$-dimensional hyperrectangular domain (box) $[a_1, b_1] \times \dots \times [a_d, b_d]$ at origin $\vec{r}_0 = (a_1, \dots, a_d)$ with edge lengths $L_i = b_i - a_i$, $i = 1, \dots, d$.
We take the Fourier amplitudes~$a_{\vec{k}}$ to be independent circular Gaussian complex random variables with zero mean and variance $s^2$ (to be fixed later), i.e., $\real(a_{\vec{k}}) \sim \mathcal{N}(0, s^2)$ and $\imag(a_{\vec{k}}) \sim \mathcal{N}(0, s^2)$.
\Cref{eq:speckle_amplitude} thus describes random interference of plane waves.
Given a set of UV cutoffs $\Lambda_1, \dots, \Lambda_d$, the sum in \cref{eq:speckle_amplitude} extends over all discrete wave vectors $\vec{k} = (k_1, \dots, k_d)$ such that $|k_i| \le \pi \Lambda_i$, where $k_i = 2 \pi \mathinner{m_i / L_i}$ and $m_1, \dots, m_d$ are integers.
This implies that $k_i L_i / 2 \pi = -\hat{m}_i, \dots, 0, \dots \hat{m}_i$ with $\hat{m}_i = \lfloor L_i \Lambda_i / 2 \rfloor$ ($\lfloor \, \cdot \, \rfloor$ denotes the floor function), which amounts to a total number of $M_i = 2 \hat{m}_i + 1$ modes in dimension~$i$.

Using the statistical properties of the Fourier amplitudes, $\ensembleaverage{a_{\vec{k}}^* a_{\vec{k}^\prime}} = 2 s^2 \delta_{\vec{k}, \vec{k}^\prime}$ ($\delta_{\vec{k}, \vec{k}^\prime}$ denotes the Kronecker delta), one can evaluate the autocorrelation function $\ensembleaverage{A^*(\vec{r}) A(\vec{r}^\prime)} = 2 s^2 \sum_{\vec{k}} \etothe{i \vec{k} (\vec{r}^\prime - \vec{r})}$.
If $\vec{r}_{\vec{j}}$ and $\vec{r}_{\vec{j}^\prime}$ are points on a commensurate regular grid with spacing $\Delta r_i = L_i / M_i$, i.e., $r_i = a_i + j_i \Delta r_i$ and $r_i^\prime = a_i + j_i^\prime \Delta r_i$ for some tuples of integers $\vec{j}$ and $\vec{j}^\prime$ with $j_i, j_i^\prime \in \set{ 0, \dots, M_i - 1 }$, one can show that $\ensembleaverage{A^*(\vec{r}_{\vec{j}}) A(\vec{r}_{\vec{j}^\prime})} = 2 s^2 M \delta_{\vec{j}, \vec{j}^\prime}$, where $M = \prod_{i = 1}^d M_i$ is the total number of modes (grid points).
Thus, the amplitudes $A(\vec{r}_{\vec{j}})$ are uncorrelated on the discrete level.

Moreover, in the continuum limit of large box extents, one can replace summation by integration according to the rule $\sum_{\vec{k}} \to \prod_{i = 1}^d (L_i / 2 \pi) \int_{- \pi \Lambda_i}^{\pi \Lambda_i} \diff k_i$, yielding
\begin{equation}
    \ensembleaverage{A^*(\vec{r}) A(\vec{r}^\prime)} = 2 s^2 \prod_{i = 1}^d (L_i \Lambda_i) \sinc \left[ \Lambda_i (x_i - x_i^\prime) \right] \,,
\end{equation}
where $\sinc(x) = \sin(\pi x) / \pi x$.
This result is consistent with the one in the discrete case since $L_i \Lambda_i \approx M_i$ and $\sinc(0) = 1$.
It is worth noting that in the limit $\Lambda_i \to \infty$, the above expression reduces to $\ensembleaverage{A^*(\vec{r}) A(\vec{r}^\prime)} = \mathinner{2 s^2 L^d} \delta(\vec{r} - \vec{r}^\prime)$, where $L^d = \prod_{i = 1}^d L_i$ is the volume of the box and $\delta(\vec{r})$ denotes the $d$-dimensional Dirac delta function.
This means that the amplitudes $A(\vec{r})$ represent Gaussian white noise in this limit.

We now define the speckle potential as the squared modulus of the amplitude field,
\begin{equation}
    V(\vec{r}) = \abs*{A(\vec{r})}^2 \,.
\end{equation}
Physically, the amplitude field~$A(\vec{r})$ plays the role of an electric field created from random interference of light, while the potential~$V(\vec{r})$ corresponds to the intensity.
To fix the variance $s^2$ of the Fourier modes in \cref{eq:speckle_amplitude}, we demand that, on average, the speckle potential takes the value $\ensembleaverage{V(\vec{r})} = V_0$, introducing the disorder strength~$V_0$.
This implies $s = \sqrt{V_0 / 2 M}$.
By construction, the speckle potential sampled this way satisfies periodic boundary conditions, which is convenient for numerical simulations, in particular for calculating the superfluid fraction via the phase-twist method.

The autocorrelation function in the continuum limit is given by
\begin{equation}
\label{eq:app:autocorrelator}
    \ensembleaverage{V(\vec{r}) V(\vec{r}^\prime)} = V_0^2 \bigg\{ 1 + \prod_{i = 1}^d \sinc^2 [ \Lambda_i (x_i - x_i^\prime) ] \bigg\} \,,
\end{equation}
which yields the variance $\ensembleaverage{V(\vec{r})^2} - \ensembleaverage{V(\vec{r})}^2 = V_0^2$.
The above result suggests the definition of the correlation length~$\sigma_i = \Lambda_i^{-1}$ as the first zero of the autocorrelation function in each dimension.

According to the Wiener--Khinchin theorem, the power spectrum can be computed from the Fourier transform of the autocorrelation function,
\begin{equation}
\label{eq:correlator_sim}
\begin{split}
    \frac{\ensembleaverage{|V(\vec{k})|^2}}{L^d} &= \int \diff \vec{r} \, \ensembleaverage{V(\vec{r}_0) V(\vec{r}_0 + \vec{r})} \etothe{- i \vec{k} \cdot \vec{r}} \\
    &= V_0^2 \bigg\{ (2 \pi)^d \delta(\vec{k}) + \prod_{i = 1}^d \sigma_i \tri \left( \frac{k_i \sigma_i}{2 \pi} \right) \bigg\} \,,
\end{split}
\end{equation}
where $\tri(x) = \max(1 - \abs{x}, 0)$ denotes the triangular function.
\Cref{eq:app:autocorrelator,eq:correlator_sim} correspond to the results reported for $d = 2$ in \cref{eq:C_x,eq:C} of the main text, respectively.

The speckle potential is thus characterized by two parameters, the scalar disorder strength~$V_0$ and the set of correlation lengths~$\vec{\sigma} = (\sigma_1, \dots, \sigma_d)$.
Isotropic disorder is characterized by the same correlation length in all dimensions, whereas stripe-like disorder is obtained if all but one correlation lengths are equal to (or exceed) the box size.

\section{Perturbation theory}
\label{app:PT}

In this appendix, we present the perturbation theory to quadratic order in the disorder strength $V_0$. These results have appeared previously in the literature~\cite{Giorgini1994,Falco2007, Gaul2011,Astrakharchik2013,Gaul2014,Cappellaro2019}, but we collect the relevant results here for clarity. In particular, we wish to evaluate the chemical potential~$\mu$, the inverse compressibility~$\kappa^{-1}$, the superfluid fraction tensor~$f_s^{i,j}$, and the speed-of-sound tensor~$c^2_{i,j}$ from the ground-state solution to the stationary GPE~\labelcref{eq:stationaryGPE}, $\Psi(\vec{r},t) = \Psi(\vec{r}) \etothe{-i \mu t/\hbar}$. 

For the purposes of perturbation theory, it is convenient to work in momentum space. There, the GPE becomes:
\begin{align}
\begin{split}
    0 =\sum_{\vec{l}}&\Big[ (\epsilon_{\vec{k}}-\mu)\delta_{\vec{k},\vec{l}}+ V(\vec{k}-\vec{l}) \\
    &+ \frac{g}{L^d}\sum_{\vec{p}} \Psi^*(\vec{p}-\vec{k})\Psi(\vec{p}-\vec{l}) \Big] \Psi(\vec{l}) \,
    \label{eq:A1:GPE_kspace}
\end{split}
\end{align}
with $\epsilon_{\vec{k}} = \hbar^2\vec{k}^2/2m$ and $L^d$ is the volume.
We next expand the condensate wave function in terms of the disorder strength:
\begin{equation}
    \Psi(\vec{k}) = \Psi^{(0)}(\vec{k}) + \Psi^{(1)}(\vec{k}) + \Psi^{(2)}(\vec{k}) + \bigO \left[ (V_0 / gn)^3 \right] \,.
\end{equation}
Substituting this expansion into \cref{eq:A1:GPE_kspace} gives
\begin{subequations}
\begin{align}
    \Psi^{(0)}(\vec{k}) &= \delta_{\vec{k},0} \sqrt{\frac{L^d\mu}{g}} \,, \\
    \Psi^{(1)}(\vec{k}) &= -\Psi^{(0)}(\vec{k}) \ \tilde{V}(\vec{k}) \,, \\
    \Psi^{(2)}(\vec{k}) &= - \Psi^{(0)}(\vec{k}) \ \frac{1}{L^d}\sum_{\vec{l}} \frac{\mu - \epsilon_{\vec{k}-\vec{l}}}{\epsilon_{\vec{k}+2\mu}} \tilde{V}(\vec{k}-\vec{l}) \tilde{V}(\vec{l}) \,.
\end{align}
\end{subequations}
At zeroth order in $V_0 / gn$, the solution to the GPE is the uniform condensate with density~$n$, while the first-order and second-order corrections depend on $\tilde{V}(\vec{k}) = V(\vec{k})/(\epsilon_{\vec{k}}+2\mu)$.

Given the solution to the wave function, we can calculate physical observables. First, we calculate the chemical potential by evaluating the density:
\begin{equation}
    n = \frac{1}{L^d} \sum_{\vec{k}}\Psi^*(\vec{k})\Psi(\vec{k}) \,.
\end{equation}
This procedure gives
\begin{equation}
\label{eq:A1:chemical_potential}
    \mu = gn + \frac{V(\vec{k} = 0)}{L^d} - \frac{1}{L^d}\sum_{\vec{k}} \frac{\left|V(\vec{k})\right|^2}{L^d} \frac{\epsilon_{\vec{k}}^3}{E_{\vec{k}}^4} \,,
\end{equation}
where $E_{\vec{k}}^2 = \epsilon_{\vec{k}}^2 + 2gn \epsilon_{\vec{k}}$ is the Bogoliubov dispersion of quasiparticles.
Using the thermodynamic definition of the compressibility, $\kappa^{-1} = n^2\partial \mu/\partial n$, we arrive at:
\begin{equation}
\label{eq:A1:compressibility}
    \kappa^{-1} = gn^2 \bigg( 1 + \frac{4}{L^d}\sum_{\vec{k}} \frac{\left|V(\vec{k})\right|^2}{L^d} \frac{\epsilon_{\vec{k}}^4}{E_{\vec{k}}^6} \bigg) \,.
\end{equation}

In order to calculate the superfluid fraction, we employ the phase-twist method, as discussed in the main text: $\Psi(\vec{r}) \to \etothe{i m \vec{v}_s \cdot \vec{r} / \hbar}\Psi(\vec{r})$, where $\vec{v}_s$ is the superfluid velocity. This amounts to boosting the momenta $\hbar \vec{k} \rightarrow \hbar \vec{k}- m\vec{v}_s$ in the momentum-space GPE~\labelcref{eq:A1:GPE_kspace}. The superfluid fraction can then be evaluated by examining the current density to linear order in $\vec{v}_s$,
\begin{equation}
    \vec{J}^i = n f_s^{i,j} \vec{v}_s^j = \frac{\hbar}{2mi} \left[ \Psi^*(\vec{r}) \nabla_{\vec{r}}^i \Psi(\vec{r}) - \Psi(\vec{r}) \nabla_{\vec{r}}^i \Psi^*(\vec{r}) \right] \,,
\end{equation}
where $i,j =x,y,z$ and $f_s^{i,j}$ is the superfluid fraction tensor. The calculation is identical to the previous case, resulting in
\begin{equation}
\label{eq:A1:superfluid_density}
    f_s^{i,j} = \delta_{i,j} - \frac{4}{L^d}\sum_{\vec{k}} \frac{\left|V(\vec{k})\right|^2}{L^d} \frac{\epsilon_{\vec{k}}^2}{E_{\vec{k}}^4} \hat{k}_i \hat{k}_j \,.
\end{equation}

Finally, we report the perturbation theory for the speed of sound. Although this is a dynamical property, we can relate it to the inverse compressibility and superfluid fraction by invoking hydrodynamics. In the hydrodynamic picture, the system obeys:
\begin{subequations}
\label{eq:A1:hydrodynamics}
\begin{align}
0 &= \mathinner{m} \partial_t v_{s,i}(\vec{r},t) +\nabla_i \left[ \mu(\vec{r},t) + \frac{1}{2} m \vec{v}_s^2(\vec{r},t) \right] \,, \\
0 &= \partial_t n(\vec{r},t) + \nabla \cdot \vec{J}(\vec{r},t) \,.
\end{align}
\end{subequations}
The first equation describes the variation of the superfluid velocity, while the second equation is the continuity equation with a current due to the superfluid flow: $\vec{J}(\vec{r},t) = f_s n (\vec{r},t) \vec{v}_s(\vec{r},t)$. We then consider small fluctuations around equilibrium, that is, we write $n(\vec{r},t) \approx n + \delta n(\vec{r},t)$. If one expands the hydrodynamic equations~\labelcref{eq:A1:hydrodynamics} to quadractic order in $\delta n(\vec{r},t)$ and $\vec{v}_s(\vec{r},t)$, one obtains a single equation for the density fluctuations:
\begin{equation}
\label{eq:A1:linearized hydrodynamics}
0 = \partial_t^2 \delta n(\vec{r},t) - \frac{f_s^{i,j} (n \kappa)^{-1}}{m} \nabla_i \nabla_j \delta n(\vec{r},t) \,.
\end{equation}
\Cref{eq:A1:linearized hydrodynamics} describes the propogation of sound modes with (squared) velocity
\begin{equation}
    m c^2_{i,j} = f_s^{i,j} (n \kappa)^{-1} \,.
\end{equation}
This is the fundamental hydrodynamic relationship~\labelcref{cHD} between the sound velocity, superfluid density, and compressibility.
Combining the results for the inverse compressibility, \cref{eq:A1:compressibility}, and the superfluid fraction, \cref{eq:A1:superfluid_density}, finally gives the desired perturbative result for the squared speed of sound:
\begin{align}
\label{eq:A1:speed_of_sound}
\begin{split}
    \frac{c^2_{i,j}}{c_0^2}= \delta_{i,j} &+ \delta_{i,j}\frac{4}{L^d}\sum_{\vec{k}} \frac{\left|V(\vec{k})\right|^2}{L^d} \frac{\epsilon_{\vec{k}}^4}{E_{\vec{k}}^6} \\
    &- \frac{4}{L^d}\sum_{\vec{k}} \frac{\left|V(\vec{k})\right|^2}{L^d} \frac{\epsilon_{\vec{k}}^2}{E_{\vec{k}}^4} \hat{k}_i \hat{k}_j \,.
\end{split}
\end{align}
\Cref{eq:chemical_potential,eq:compressibility,eq:superfluid_density,eq:speed_of_sound} reported in the main text follow by specializing the results derived in this appendix to two spatial dimensions and upon taking the ensemble average.

\section{Damping of the sound mode}
\label{app:damping}

Next, we obtain the damping of the sound mode following the method of quantum hydrodynamics employed in Ref.~\cite{Giorgini1994}. In this approach, we perturbatively evaluate the density--density response function in the imaginary-time formalism~\cite{Bruus2004}. This is equivalent to investigating the Bogoliubov modes on top of a disordered condensate~\cite{Gaul2011}.

The starting point is to assume that the long-wavelength and low-frequency behaviour of the structure factor is dominated by the sound mode. Identically, we assume that the density--density response function has a pole at the corresponding sound mode. For an anisotropic sound mode, this implies a density--density response function of the form
\begin{align}
\label{eq:A2:density_density_bare}
    &D^0_{n,n}(\vec{q},i\omega_n) \nonumber \\
    &\qquad=  \int \diff \vec{r} \, \etothe{-i \vec{q} \cdot \vec{r}} \int_0^{1/T} \diff \tau \, \etothe{i \omega_n \tau} \quantumaverage*{\mathcal{T}_{\tau} \delta n(\vec{r},\tau) \delta n(0,0)} \nonumber \\
    &\qquad\approx \frac{\vec{q} \cdot \vec{n}_s \cdot \vec{q}}{(i\omega_n)^2 - \vec{q} \cdot \vec{c}^2 \cdot \vec{q}}
\end{align}
in the long-wavelength and low-frequency limit.
In the above expression, $\mathcal{T}_{\tau}$ is the imaginary-time-ordering operator, $T$ is the temperature (which we send to zero), and $\delta n(\vec{r},\tau) = n(\vec{r},\tau)-n_0$ are the fluctuations around the background density~$n_0$.
The form of \cref{eq:A2:density_density_bare} reduces to the isotropic expression in Ref.~\cite{Giorgini1994} for isotropic disorder, and is consistent with the compressibility sum rule, see \cref{kappa}.

The leading mechanism for the damping is coupling between density and local phase fluctuations, $\theta(\vec{r},t)$, which arises naturally in quantum hydrodynamics:
\begin{equation}
U_{\text{pert}} = \int \diff \vec{r} \, \frac{1}{2}\delta n(\vec{r},\tau) \left[ \nabla_{\vec{r}} \theta(\vec{r},\tau) \right]^2 \,.
\label{eq:A2:perturbation}
\end{equation}
We determine the damping by calculating the imaginary part of the leading self-energy correction to the density--density propagator from \cref{eq:A2:perturbation}:
\begin{multline}
\label{eq:sigma}
    \Sigma(\vec{q},i\omega_n) = \left(\frac{i\omega_n}{n_0 \vec{q}^2}\right)^2 \frac{1}{L^d} \sum_{\vec{k}} \frac{|V(\vec{k}-\vec{q})|^2}{L^d} \left(\vec{q} \cdot \vec{k}\right)^2 \\
    \times \left(\frac{n_0 (\vec{k}-\vec{q})^2}{E_{\vec{k}-\vec{q}}^2}\right)^2  \frac{E_{\vec{k}}^2}{n_0 \vec{k}^2} \frac{1}{(i\omega_n)^2 - E_{\vec{k}}^2} \,.
\end{multline}
The total density--density response function then has the form
\begin{multline}
    \left(\vec{q} \cdot \vec{n}_s \cdot \vec{q}\right) D_r^{-1}(\vec{q},i\omega_n) \\
    \approx (\omega - \vec{q} \cdot \vec{n}_s \cdot \vec{q}) -i n_0 q^2 \imag\left[\Sigma(\vec{q},i\omega_n \approx c_0 q)\right] \,,
\end{multline}
where we have neglected the real part of the self-energy, which is included in the sound pole, and have placed the self-energy on-shell: $i \omega_n \approx c_0 q + \bigO(V_0/gn)$. Consequentially, the structure factor becomes a Lorentzian,
\begin{equation}
    S(\vec{q},\omega) = \frac{\vec{q} \cdot \vec{n}_s \cdot \vec{q}}{\pi}\frac{2 c_0 q \gamma_{\vec{q}}}{(\omega^2 - \vec{q} \cdot \vec{n}_s \cdot \vec{q})^2+ 4(c_0 q)^2 \gamma_{\vec{q}}^2} \,.
\end{equation}
The width of the Lorentzian is simply the damping rate of the phonons, for which 
we find the result
\begin{equation}
\label{eq:A2:damping_rate_discrete}
    \frac{\gamma_{\vec{q}}}{c_0 q} = \frac{1}{2} \frac{\pi}{2}  \frac{q}{m^2c_0^4} 
    \frac{1}{L^d} \sum_{\vec{k}} \frac{\abs[\big]{V(\vec{k}) }^2}{L^d} \big( \vec{\hat{q}} \cdot \vec{\hat{k}} \big)^2 \delta(q-k) \,.
\end{equation}
\Cref{eq:A2:damping_rate_discrete} contains a Dirac delta function and should formally be evaluated after taking the continuum limit. This procedure yields the result for the damping rate in $d$ spatial dimensions:
\begin{equation}
\label{eq:A2:damping_rate}
    \frac{\gamma_{\vec{q}}}{c_0 q} = \frac{1}{2} \frac{\pi}{2} \frac{\Omega_d}{(2 \pi)^d} \frac{q^d}{m^2c_0^4} 
    \int \frac{\diff \Omega_{\vec{k}}}{\Omega_d} \frac{\abs[\big]{V \big( q \vec{\hat{k}} \big)}^2}{L^d} \big( \vec{\hat{q}} \cdot \vec{\hat{k}} \big)^2 \,,
\end{equation}
where $\Omega_d$ is the solid angle of a $d$-dimensional sphere, while $\vec{\hat{k}}$ and $\vec{\hat{q}}$ are unit vectors in the $\vec{k}$ and $\vec{q}$ direction, respectively.
Finally, the result reported in \cref{eq:damping_rate} of the main text follows for $d = 2$ after ensemble-averaging.

\section{Finite-size effects}
\label{app:finite_size}

In this section, we discuss some of the nuances in extracting the speed of sound and its damping in our simulations.
In particular, we discuss the dynamics of single realizations of disorder and how it can differ from the average due to finite-size effects.
Our protocol is the one described in \cref{sec:TM}: we prepare the system in the ground state subject to both the disorder potential and the weak sinusoidal probe potential in \cref{eq:V_probe}.
At time $t=0$, the probe potential is removed and we track the dynamics by evaluating $\quantumaverage{\sin(\vec{q} \cdot \vec{r})}(t)$.

\subsection{Mode-coupling effects in single realizations: beating versus damping}

\begin{figure}[htb]
	\includegraphics[width=\columnwidth]{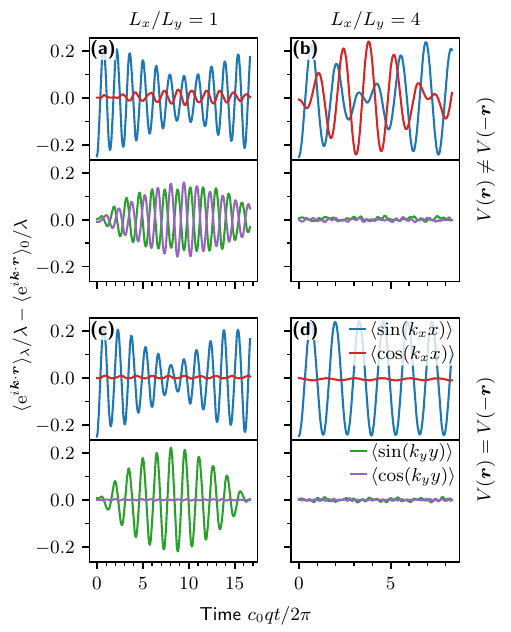}%
    \subfloat{\label{fig:beating:a}}%
    \subfloat{\label{fig:beating:b}}%
    \subfloat{\label{fig:beating:c}}%
    \subfloat{\label{fig:beating:d}}%
	\caption{\label{fig:beating}%
        Finite-size beating effects of long-wavelength phonons.
        A phonon with wavelength equal to the box size~$L_x$ is excited in the $x$~direction (wave vector $\vec{q} = \vec{\hat{e}}_x \mathinner{2 \pi / L_x}$), subject to isotropic 2D disorder with strength~$V_0 / gn = 8$ and correlation length $\sigma_x = \sigma_y = 2 \sqrt{2} \xi$.
        (a)~In a square box ($L_x / L_y = 1$) and for a disorder potential without parity symmetry [$V(\vec{r}) \neq V(-\vec{r})$], one can observe a strong response of the four observables $\quantumaverage{\sin(k_x x)}$, $\quantumaverage{\cos(k_x x)}$, $\quantumaverage{\sin(k_y y)}$, and $\quantumaverage{\cos(k_y y)}$ with $k_x = 2 \pi / L_x$ and $k_y = 2 \pi / L_y$.
        (b)~In a rectangular box ($L_x / L_y = 4$ with fixed volume $L_x L_y$), the coupling to the $y$~direction disappears.
        (c)~If parity symmetry [$V(\vec{r}) = V(-\vec{r})$] is imposed on the disorder potential, only the observables $\quantumaverage{\sin(k_x x)}$ and $\quantumaverage{\sin(k_y y)}$ are strongly excited.
        (d)~The combination of (b) and (c), i.e., a parity-symmetric disorder potential in a rectangular box, yields a single dominant excitation where only $\quantumaverage{\sin(k_x x)}$ oscillates at a single frequency.%
    }
\end{figure}

At the level of single realizations, we can observe dynamics that is not necessarily well described by the damped oscillation in \cref{eq:sound_fit}.
Depending on the system geometry, the phonon wavelength, and the characteristics of the disorder potential, the number of strongly coupled momentum modes can vary significantly and result in different dynamical behavior.

In the regime where the energy of the collective excitation is spread over a limited number modes, the observable $\quantumaverage*{\sin(\vec{q} \cdot \vec{r})}(t)$ typically exhibits a beating of few dominant frequencies with revivals of high contrast and no visible damping on the investigated time scales.
This is the case for stripe-like 1D disorder and for the low-$q$ sound modes in 2D disorder.

As an example, we investigate in \cref{fig:beating} the scenario of isotropic 2D disorder where a phonon is excited at the wavelength of the box size in the disorder direction (wave vector $\vec{q} = \vec{\hat{e}_x} \mathinner{2 \pi / L}$).
In a square box, typical realizations then exhibit a strong response in the observables $\quantumaverage{\cos(\vec{k} \cdot \vec{r})}(t)$ and $\quantumaverage{\sin(\vec{k} \cdot \vec{r})}(t)$ for both $\vec{k} = \vec{\hat{e}_x} \mathinner{2 \pi / L}$ and $\vec{k} = \vec{\hat{e}_y} \mathinner{2 \pi / L}$, as shown in \cref{fig:beating:a}.
In this situation, the four lowest-lying momentum modes in the $x$ and $y$ direction at $\vec{k} = \pm q \vec{\hat{e}_x}$ and $\vec{k} = \pm q \vec{\hat{e}_y}$, respectively, are predominantly excited.
The strong coupling to the modes along $y$ is a consequence of the square geometry of the box, causing momentum modes of the same magnitude to be close in energy.
We have checked that in a rectangular box, elongated in the $x$~direction, the coupling to the $y$~direction is significantly reduced, see \cref{fig:beating:b}.
In addition, since the speckle disorder potential is generally not parity symmetric [$V(\vec{r}) \neq V(-\vec{r})$], neither is the dynamic structure factor [$S(\vec{q}, \omega) \neq S(-\vec{q}, \omega)$], resulting in a coupling between the $+\vec{k}$ and $-\vec{k}$ modes that manifests itself as a beating.
We have checked that by imposing parity symmetry on the sampled speckle potential (which can be achieved by choosing the Fourier amplitudes in \cref{eq:speckle_amplitude} to be real and adjusting the normalization appropriately), the signal with opposite parity of the probe potential, $\quantumaverage{\cos(\vec{k} \cdot \vec{r})}(t)$, becomes significantly weaker, see \cref{fig:beating:c}.
Consequently, for a parity-symmetric disorder potential in a rectangular box, we find only a single dominant excitation of the observable $\quantumaverage{\sin(\vec{k} \cdot \vec{r})}(t)$ for $\vec{k} = \vec{q}$, which oscillates at a single frequency, as demonstrated in \cref{fig:beating:d}.

As the number of strongly coupled modes increases, single realizations typically begin to exhibit exponential damping akin to \cref{eq:sound_fit}.
This can most clearly be observed for 2D disorder of sufficient strength at larger phonon wave numbers~$q$, see \cref{fig:damping_2d:b}.
It is this regime we consider in perturbation theory, see \cref{eq:damping_rate}, i.e., the continuum limit where there is a macroscopic number of coupled $\vec{k}$ modes.
The beating described above can therefore be regarded as a finite-size effect, while damping is the generic behavior expected for systems in the thermodynamic limit.

Finally, it should be mentioned that mode coupling effects leading to beating or damping can also arise due to nonlinear effects even in stripe-like 1D disorder for phonons propagating along the stripes in the $y$~direction.

\subsection{Single realizations versus disorder-averaging}

The sound velocities~$c$ and damping rates~$\gamma_{\vec{q}}$ reported in \cref{fig:superfluidity_2d,fig:damping_2d,fig:superfluidity_1d,fig:damping_1d} in the main text have been extracted from the ensemble-averaged observable $\ensembleaverage{\quantumaverage{\sin(\vec{q} \cdot \vec{r})}}(t)$ by fitting the exponentially damped oscillation in \cref{eq:sound_fit} to the data.
In some sense, the act of disorder-averaging can be thought of as mimicking a larger set of modes coupled by the disorder potential, revealing the damped oscillations expected from mode coupling in the continuum limit.

In order to probe the speed of sound in our simulations, we have excited phonons with wavelengths equal to the box size ($q = 2 \pi / L$) and examined the resulting sound-wave oscillations via the observable $\quantumaverage{\sin(\vec{q} \cdot \vec{r})}(t)$.
Due to the finite-size effects described above, such long-wavelength phonons typically exhibit beat notes in single realizations, which can be approximated by a sum of sinusoids.
The speed of sound may therefore alternatively be obtained by extracting the dominant frequencies from each realization and averaging over the resulting set of frequencies.
We have checked that in the regime where the quality factor is well above one, extracting the speed of sound either from the ensemble average or from individual realizations yields consistent results.

Moreover, the procedure of averaging the observable $\quantumaverage{\sin(\vec{q} \cdot \vec{r})}(t)$ allows us to assign a damping rate even to those configurations where, due to finite-size effects, single realizations exhibit beating instead of damping, e.g., for stripe-like 1D disorder, as discussed above.
Remarkably, as can be seen in \cref{fig:damping_1d}, the resulting damping rates are consistent with the predictions from perturbation theory, although we note that the damping on average is due to dephased beating in this case.
By contrast, for isotropic 2D disorder, even single realizations exhibit damped oscillations (except for phonons with wavelength equal to the box size), see \cref{fig:damping_2d}.
We have checked that, in the cases where damping rates can be extracted from individual realizations in a meaningful way, they agree on average with the damping rate obtained from the ensemble-averaged signal up to statistical errors.

\subsection{Damping versus dephasing}

\begin{figure}[htb]
	\includegraphics[width=\columnwidth]{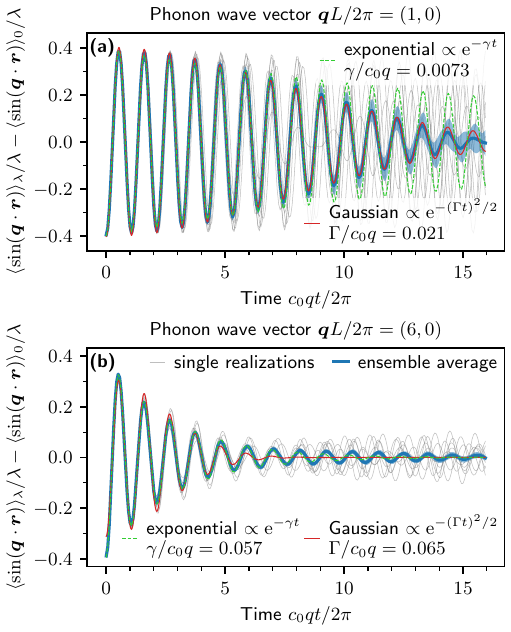}%
    \subfloat{\label{fig:single_versus_average:a}}%
    \subfloat{\label{fig:single_versus_average:b}}%
	\caption{\label{fig:single_versus_average}%
        Damping versus dephasing for isotropic 2D disorder with $V_0 / gn = 2$ and $\sigma_x = \sigma_y = 2 \sqrt{2} \xi$.
        The thin gray lines depict the dephasing of individual realizations, yielding suppressed revivals in the ensemble average (blue line).
        The green dashed and solid red lines show fits of decaying oscillations with exponential and Gaussian envelopes, respectively.
        For a phonon wave vector of $\vec{q} L / 2 \pi = (1, 0)$, the dephased signal is better described by a Gaussian envelope (a), while for $\vec{q} L / 2 \pi = (6, 0)$, damping with exponential envelope dominates (b).%
    }
\end{figure}

In our simulations, we observe damping on average if the dynamics is dominated by mode coupling and revivals occurring at later time scales are suppressed via dephasing.
However, in certain regimes, e.g., when disorder and thus mode coupling is weak, the dynamics can also be dominated by dephasing.

To further investigate the interplay between damping and dephasing, we consider the effect of taking the ensemble average of the observable $\quantumaverage{\sin(\vec{q} \cdot \vec{r})}(t)$.
If this quantity contains oscillating terms $\propto \cos(\omega_{\vec{q}} t)$, their ensemble average is given by $\ensembleaverage{\cos(\omega_{\vec{q}} t)} = \int \diff \omega \, p_{\vec{q}}(\omega) \cos(\omega t)$.
Here, $p_{\vec{q}}(\omega)$ is the probability distribution of phonon frequencies, and we have neglected correlations with other random quantities, e.g., amplitudes.
If $p_{\vec{q}}(\omega)$ can be approximated by a Gaussian with mean $\ensembleaverage{\omega}_{\vec{q}}$ and variance $\Gamma_{\vec{q}}^2$, ensemble-averaging results in dephasing with a Gaussian envelope, $\ensembleaverage{\cos(\omega_{\vec{q}} t)} = \etothe{- (\Gamma_{\vec{q}} t)^2 / 2} \cos(\ensembleaverage{\omega}_{\vec{q}} t)$.

In \cref{fig:single_versus_average}, we compare the situation where the amplitude of the ensemble-averaged signal $\ensembleaverage{\quantumaverage{\sin(\vec{q} \cdot \vec{r})}}(t)$ decreases either with an exponential or a Gaussian envelope.
The strength of the isotropic 2D disorder potential is chosen to be moderate, $V_0 / gn = 2$, such that damping is expected to be small at the lowest phonon wave vector $\vec{q} L / 2 \pi = (1, 0)$, see \cref{fig:single_versus_average:a}.
In this case, the dynamics is dominated by dephasing and a Gaussian envelope fits the decay of the ensemble-averaged signal better.
By contrast, for the larger phonon wave vector $\vec{q} L / 2 \pi = (6, 0)$ in \cref{fig:single_versus_average:b}, single realizations exhibit sizable damping and the ensemble average is better described by a damped oscillation with exponential envelope.

\section{Impact of long-wavelength perturbations on the validity of hydrodynamics}
\label{app:infrared_cutoff}

\begin{figure}[htb]
	\includegraphics[width=\columnwidth]{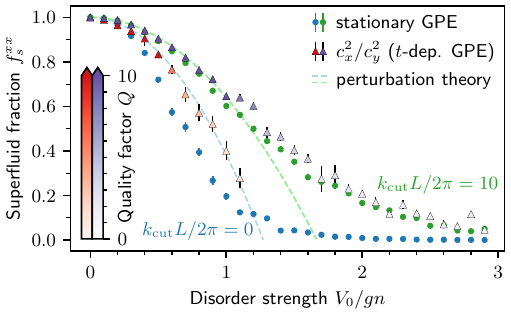}%
	\caption{\label{fig:superfluid_fraction_filter}%
        Impact of long-wavelength perturbations on the validity of hydrodynamics.
        The plot shows the superfluid fraction~$f_s^{xx}$ as a function of the disorder strength~$V_0$ for stripe-like 1D disorder with $\sigma_x = 2 \sqrt{2} \xi$ in the absence (blue/red colors) and presence (green/purple colors) of a high-pass filter, removing Fourier components of the disorder potential below a certain cutoff wave number~$k_{\mathrm{cut}}$.
        The validity of the hydrodynamic relation~\labelcref{cHD} is assessed by comparing the superfluid fraction computed from the Leggett integral over the ground-state density profile (circles) to the value obtained from the ratio of the squared sound velocities extracted from time-dependent GPE simulations (triangles).
        The color scale shows the quality factor~$Q$ of the damped oscillations after exciting a phonon with wave vector $\vec{q} = \vec{\hat{e}_x} \mathinner{2 \pi / L}$ in the disorder direction.
        The dashed lines depict the predictions from perturbation theory according to \cref{eq:superfluid_density}.
        In the presence of the high-pass filter ($k_{\mathrm{cut}} L / 2 \pi = 10$), the hydrodynamic description remains accurate over an extended range of disorder strengths.%
    }
\end{figure}

In this appendix, we investigate the influence of long-wavelength Fourier components in the disorder potential on the validity of the hydrodynamic relation~\labelcref{cHD}.
Generally, hydrodynamics describes the salient dynamics of a physical system in terms of a few macroscopic variables that are insensitive to the underlying microscopic details.
In our disorder context, we thus expect the hydrodynamic description of sound propagation at long wavelengths to be accurate as long as the system is not strongly affected by the disorder on the relevant length scales.

As shown by \cref{eq:compressibility,eq:superfluid_density,eq:speed_of_sound}, the impact of disorder on the compressibility, superfluid fraction, and speed of sound is to leading order determined by the disorder power spectrum~$\ensembleaverage{\abs{V(\vec{k})}^2} / L^2 \propto V_0^2 \sigma_x \sigma_y$.
In \cref{fig:superfluidity_2d,fig:superfluidity_1d}, it can be seen that the hydrodynamic relation fails if the strength~$V_0$ and correlation lengths $(\sigma_x, \sigma_y)$ of the disorder potential become too strong.
In particular, for stripe-like 1D disorder ($\sigma_y = L$), the spectral power is enhanced at wave vectors $\vec{k} = k_x \vec{\hat{e}}_x$ along the disorder direction with respect to 2D disorder ($\sigma_y < L$).
This explains the overall stronger response of the system to 1D disorder for a fixed value of $V_0$.
In addition, the spectral power is largest at the longest wavelengths commensurate with the system size, see \cref{fig:speckle_disorder:a}).
In \cref{fig:superfluid_fraction_filter}, we show that by removing these long-wavelength Fourier components using a high-pass filter, $V(\vec{k}) \to \theta(k - k_{\mathrm{cut}}) V(\vec{k})$ with cutoff wave number~$k_{\mathrm{cut}}$, the validity of the hydrodynamic relation~\labelcref{cHD} improves substantially.
Importantly, such a filter procedure is not merely a theoretical tool to characterize the system's response to synthetic external perturbations, but can also be employed experimentally through established potential-shaping techniques, e.g., using digital micromirror devices~\cite{Chauveau2023}.

\bibliography{references}

\end{document}